\documentclass[preprint,showpacs,preprintnumbers,amsmath,amssymb]{revtex4}

\usepackage{graphicx}
\usepackage{dcolumn}
\usepackage{bm}

\usepackage{color}
\RequirePackage[colorlinks=true,citecolor=blue,linkcolor=blue,breaklinks]{hyperref}

\newcommand{\beq}{\begin{equation}}
\newcommand{\eeq}{\end{equation}}
\newcommand{\mbf}{\mathbf}
\newcommand{\mcl}{\mathcal}
\newcommand{\brr}{\begin{array}}
\newcommand{\err}{\end{array}}
\newcommand{\equref}{Eq.\,\eqref}

\newcommand{\bq}{\begin{eqnarray}}
\newcommand{\eq}{\end{eqnarray}}

\newcommand{\dcs}{differential cross section \,}
\newcommand{\ina}{invariant amplitude \,}
\newcommand{\frma}{four-momenta \,}
\newcommand{\spp}{strange particle production \,}

\begin{document}

%\preprint{}

\title{Strange Particle Production Via The Weak Interaction}

\author{G.~B.~Adera}
  \email{adera@sun.ac.za} 
   \affiliation{Department of Physics, Stellenbosch University, Stellenbosch 7600, South Africa}
\author{B.~I.~S.~Van Der Ventel}
  \email{bventel@sun.ac.za}
  \affiliation{Department of Physics, Stellenbosch University, Stellenbosch 7600, South Africa}
\author{D.~D.~van Niekerk}
  \affiliation{Department of Physics, Stellenbosch University, Stellenbosch 7600, South Africa}
\author{T.~Mart}
   \affiliation{Departemen Fisika, FMIPA, Universitas Indonesia, Depok 16424, Indonesia}

\date{\today}

\begin{abstract}
The differential cross sections for the neutrino-induced weak charged current production of strange particles in the threshold energy region are presented. The general representation of the weak hadronic current is newly developed in terms of eighteen unknown invariant amplitudes to parametrize the hadron vertex. The Born term approximation is used for the numerical calculations in the framework of the Cabibbo theory and $\rm{SU(3)}$ symmetry. For unpolarized octet baryons four processes are investigated, whereas in the case of polarized baryons only one process is chosen to study the sensitivity of the \dcs to the various polarizations of the initial state nucleon and the final state hyperon. 
\end{abstract}


\maketitle
\section{\label{sec:level1}Introduction}

The study of neutrino-induced weak interactions has become one of the frontiers of theoretical and experimental research in the fields of cosmology, astrophysics, particle physics, and nuclear physics. For instance, it allows for the analysis of the various oscillation experiments, the detailed study of the strange-quark content of the nucleon, the investigation of the structure of the hadronic weak current, and the estimation of the atmospheric neutrino backgrounds for nucleon decay searches \cite{NiSo05,ASu06}. 

Neutrinos are very important in the study of the strange-quark contribution to the nucleon spin. In the near future experiments such as Miner$\nu$a will allow physicists to gain considerable insight regarding the structure of the nucleon and the hadronic weak current via the neutrino-induced weak production of strange particles \cite{NiSo05}. 

After the experimental evidence was reported by Ref. \cite{SJB74}, the first extensive theoretical studies of the strange particle productions via the weak interaction in comparison with experiment were done in Refs. \cite{RSH75,ME76,DE80}. Shrock and Mecklenburg independently studied the associated production of charged current (CC) reactions by employing the Cabibbo theory with $\rm{SU(3)}$ symmetry and neutral current (NC) in the framework of the Weinberg-Salam model, whereas Dewan focused on the CC and strangeness changing ($\Delta S = 1$) \spp reactions. They decomposed the norm squared invariant matrix element in terms of the helicity amplitudes which would be determined from the Born diagram.

The main purpose of this paper is the investigation of strange particle production via the weak interaction of the neutrino and nucleon near threshold energy. We focus on the following specific CC reactions: 
\bq
\label{eq:ch1_0a}
\nu\text{p} \rightarrow \mu^-[\text{K}^+\Sigma^+]:
\,\,\,\,\text{CC1}\,\,\,\,\,\,\,\,\,\,\,\,\,\,\,\,\,\,\,\,\,\,\,\,\,\,\,\,\,\,\,\,\,\,\,\,\,\,\,\,\,\,\,\,\,\,\,\,\,\,\,\,\,\,\,\,\,\,\,\,\,\,\,\,\,\,\,\, \\
\label{eq:ch1_0b}
\nu\text{n} \rightarrow \mu^-[\text{K}^0\Sigma^+,\,\text{K}^+\Sigma^0,\,\text{K}^+\Lambda]: \,\,\,\,\text{CC2},\,\,\,\text{CC3},\,\,\,\text{CC4},
\eq

which are the neutrino-induced associated productions of octet baryons and pseudoscalar mesons. What makes these reactions interesting is not only that they provide a  possibility to test the Cabibbo V-A theory and the $\rm{SU(3)}$ symmetry but also that they allow the investigation of the sensitivity of the differential cross section to the various baryon polarizations owing to the experimental feasibility to measure polarization of hyperons \cite{RSH75}.

The general formalism is done in the relativistic framework in the rest frame of the nucleon. The kinematic part of the reactions is specified in two planes: the leptonic and the hadronic planes. Unlike the leptonic transition current, there are no well developed and tested gauge theories that allow us to calculate the hadronic weak transition current. Therefore, we have developed a new model-independent approach to evaluate the invariant matrix element. Hence we derive, for the first time, the most general representation of the hadronic weak current operator for strange particle production.

For the numerical calculation of the \dcs for CC associated productions, we follow a  scheme similar to that employed by Ref. \cite{ME76} for evaluating the Born diagram that approximates the hadronic vertex. That is, we apply the Cabibbo V-A theory in relation to the conserved vector current (CVC) hypothesis as well as the $\rm{SU(3)}$ predictions of the strong coupling constants to determine the invariant amplitudes of the weak hadronic currents.

\section{\label{sec:level2}Formalism}
In this section we present the relativistic formalism for the neutrino-induced charged-current production of the strange particles. The \dcs will be constructed as a contraction between a leptonic tensor and a hadronic tensor. The electroweak theory of Glashow, Salam and Weinberg is used to calculate the leptonic tensor. The hadronic current is determined from the newly derived general form of the weak hadronic current which is expressed in terms of eighteen invariant amplitudes that parametrize the hadron vertex. In order to extract the values of the eighteen structure functions we introduce the Born model.

\subsection{\label{sec:ch2p1}Differential cross section}
\begin{figure}[h]
\centering 
\includegraphics[scale=0.825]{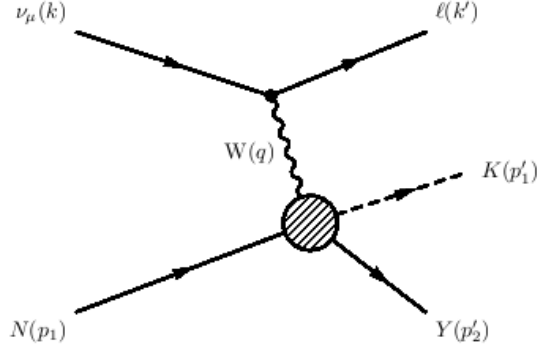} 
  \caption{The lowest order Feynman diagram of neutrino-induced strange particle production.} 
\label{fig:ch2_0}
\end{figure}

The processes under consideration are of the form 
\begin{equation}
\nu(k, h) + \text{N}(p_1, s_1) \rightarrow \ell(k', h') + \text{K}(p'_1) + \text{Y}(p'_2, s'_2).
\label{eq:ch2_a} 
\end{equation}
Here $\nu$ and $\ell$ refer to the initial neutrino and the final state lepton, respectively, $\text{N}$ and $\text{Y}$ represent the initial nucleon and the final hyperon states, respectively, and  $\text{K}$ stands for the final state pseudoscalar meson. The corresponding four-momenta labels of the particles are also given in parentheses along with the spin polarization. The lowest order Feynman diagram for the strange particle production reaction is shown in Fig.\,\ref{fig:ch2_0}.
\\

The most general form of the differential cross section of the reaction process that can be represented by Fig. \ref{fig:ch2_0} is constructed based on Fermi's Golden rule. In the rest frame of the nucleon \cite{BP04}, 

\begin{equation}
\begin{array}{ll}
\rm {d \sigma}  = & \dfrac{(2\pi)^4 \delta(k + p_1 - k' - p'_1 - p'_2)}{|\mathbf{v} - \mathbf{v_{1}}|} |\mathcal{M}|^2 \left\lbrace \dfrac{1}{(2\pi)^3} \rm d^3 \mathbf{k}' \dfrac{1}{(2\pi)^3} \dfrac{\rm d^3 \mathbf{p}'_1}{2E_{p'_1}}  \dfrac{1}{(2\pi)^3} \rm d^3 \mathbf{p}'_2\right\rbrace.
\end{array}
\label{eq:ch2_b}
\end{equation}

The lowest order Feynman diagram is used to construct the \ina for the reaction processes in question. In general the diagram in Fig. \ref{fig:ch2_0} contains four major components: the weak leptonic transition, the weak hadronic transition, the gauge boson propagator, and the interaction vertices. Thus the general expression of $\mcl{M}$ becomes:

\beq
-i\mcl{M} = \left[\bar{u}_{l}(\mbf{k}',h') \eta_l \gamma_{\mu} (I - 
\gamma_5){\nu}(\mbf{k},h)\right]iD^{\mu \nu} \left\langle \text{K}(p'_1)\text{Y}(p'_2) \left| \eta_h \hat{J}_{\nu}(q)\right|\text{N}(p_1) \right\rangle,
\label{eq:ch2_c} 
\eeq

where for convenience the neutrino's Dirac spinor is represented by ${\nu}(\mbf{k},h)$ instead of $u_{\nu}(\mbf{k},h)$ and $D^{\mu \nu}$ is the gauge boson propagator
\beq
D^{\mu \nu} = \dfrac{-g^{\mu \nu } + q^{\mu}q^{\nu}/M^2_{W}}{q^2 - M^2_{W}},
\label{eq:ch2_d}
\eeq

and the hadronic transition current can be written as 

\beq
\left\langle \text{K}(p'_1)\text{Y}(p'_2) \left|\eta_h \hat{J}_{\nu}\right|\text{N}(p_1) \right\rangle = \bar{u}_{Y}(\mbf{p}'_2, s'_2)\eta_h J_{\nu}(q)u_{N}(\mbf{p}_1,s_1)
\label{eq:ch2_e}
\eeq
with the two factors $\eta_l$ and $\eta_h$ given in Table \ref{tab:ch2_0}. Note that here $\eta_c$ is the Cabibbo factor and it may have any of the following forms in term of the Cabibbo angle $\theta_c$: \beq
\eta_c = \left\lbrace \brr{ll}
\cos \theta_c, & \text{for CC}, \Delta S = 0 \vspace*{0.1cm},\\
\sin \theta_c, & \text{for CC}, \Delta S = 1 \vspace*{0.1cm},\\
1 & \text{for NC}.
\err\right. 
\label{eq:ch2_f}
\eeq

The condition $Q^2 \ll M^2_{W}$ ($Q^2 = -q^2$) must be satisfied such that $D^{\mu \nu} \rightarrow {g^{\mu \nu}}/{M^2_{W}}$, which in turn relates the weak coupling constant with the Fermi constant $G_F = 1.166 \times 10^{-5} \,\text{GeV}^{-2}$. That is, 

\beq
\dfrac{G_F}{\sqrt{2}} = \dfrac{g^2}{8M^2_W}.
\label{eq:ch2_g}
\eeq
Then $\mcl{M}$ becomes \cite{VP06}

\beq
\mcl{M} = \dfrac{G_F}{\sqrt{2}}\eta \left[\bar{u}_{l}(\mbf{k}',h') \gamma_{\mu} (I - 
\gamma_5)\nu(\mbf{k}, h) \right] \left[ \bar{u}_{Y}(\mbf{p}'_2,s'_2) J^{\mu}(q)u_{N}(\mbf{p}_1,s_1)\right] 
\label{eq:ch2_h}
\eeq
where  
\beq
\eta = \left\lbrace  \brr{ll} \eta_c & \text{for CC}\vspace*{0.15cm} \\
\dfrac{\eta_c}{2} & \text{for NC}.
\err \right. 
\label{eq:ch2_i}
\eeq

Thus the norm squared invariant matrix element may be written as the contraction between the leptonic tensor, $L_{\mu \nu}$, and the hadronic tensor, $W^{\mu \nu}$:

\beq
|\mcl{M}|^2 = \dfrac{G^2_F \eta^2}{2} L_{\mu \nu} W^{\mu \nu},
\label{eq:ch2_j}
\eeq
where 
\beq
L_{\mu \nu} = \left[\bar{u}_{l}(\mbf{k}',h') \gamma_{\mu} (I - \gamma_5) \nu(\mbf{k}, h) \right]\left[\bar{u}_{l}(\mbf{k}',h') \gamma_{\nu} (I - \gamma_5)\nu(\mbf{k}, h) \right]^* 
\label{eq:ch2_k}
\eeq
and
\beq
W^{\mu \nu} = \left[ \bar{u}_{Y}(\mbf{p}'_2,s'_2) J^{\mu}(q)u_{N}(\mbf{p}_1,s_1)\right] \left[ \bar{u}_{Y}(\mbf{p}'_2,s'_2) J^{\nu}(q)u_{N}(\mbf{p}_1,s_1)\right]^*.\,\,\,
\label{eq:ch2_l}
\eeq

\begin{table}[h]
\caption{\label{tab:ch2_0}The expressions for $ \eta_l$ and $\eta_h$ in terms of the weak coupling constant. }
\begin{ruledtabular}
\begin{tabular}{lll}
&\,\,CC&\,\,\,\,\,\,\,\,\,NC \\ 
\hline
 $\eta_l$ & $\dfrac{-ig}{2\sqrt{2}}$ & $\dfrac{-ig}{4}\dfrac{M_Z}{M_{W^{+}}}$\vspace*{0.125cm}\\ 
$\eta_h$ & $\dfrac{-ig}{2\sqrt{2}}\eta_c$ & $\dfrac{-ig}{4}\dfrac{M_Z}{M_{W^{+}}}\eta_c$. 
\end{tabular}
\end{ruledtabular}
\end{table}

Since the study is done at the elementary level, spin-$\frac{1}{2}$ free particles are described in terms of the Dirac spinor fields with non-covariant normalization condition:
\begin{equation}
 u(\mathbf{p}, s) =  \sqrt{\dfrac{E + m}{2E}}\left( \begin{array}{c}
\phi^{s}\vspace*{0.20cm}\\
\dfrac{\vec{\sigma}\cdot \mathbf{p}}{E + m}\phi^{s} 
\end{array}\right), \,\,\,\,\,\,\,\,\,\,\,\,\,u^{\dagger}(\mathbf{p}, s)u(\mathbf{p}, s') = \delta_{ss'}.  
\label{eq:ch2_m}
\end{equation}
However, in the small-mass limit, the helicity representation of the Dirac spinor is more appropriate. That is,
\begin{equation}
u(\mbf{p}, h) = \sqrt{\dfrac{E + m}{2E}}\left(\begin{array}{cc}
\phi^{h}(\hat{\mbf{p}})\vspace{0.25cm}\\
\dfrac{h|\mathbf{p}|}{E + m}\phi^{h}(\hat{\mbf{p}})
\end{array}\right), \,\,\,\,\,\,\,\, h = \pm 1.
\label{eq:ch2_n}
\end{equation}
Thus the normalized helicity eigenstates with their corresponding eigenvalues $h = \pm 1$ are: 
\begin{equation}
 \phi^{h = +1}(\varphi,\, \theta) = \left(\begin{array}{c}
 \cos \frac{\theta}{2} \\
\sin \frac{\theta}{2} e^{i \varphi} 
\end{array}\right) \,\,\,\,  \text{and} \,\,\,\,  
\phi^{h = -1}(\varphi,\, \theta) = \left(\begin{array}{c}
 -\sin \frac{\theta}{2}e^{-i \varphi} \\
\cos \frac{\theta}{2} 
\end{array}\right).
\label{eq:ch2_o}
\end{equation}
The leptonic tensor is derived from the leptonic weak transition current by applying the Feynman trace technique. This approach avoids the explicit use of the Dirac spinors and the gamma matrices for the calculation of the differential cross section. The projection operators play a key role in the further simplification of the tensor. Thus the energy-spin projection operator is defined as 

\begin{equation}
u(\mathbf{p}, s)_{\alpha} \bar{u}(\mathbf{p}, s)_{\beta} = \left[\,\dfrac{\displaystyle{\not}p + m }{2E}\,\dfrac{I + \gamma_5 \displaystyle{\not}{s}}{2}\,\right]_{\alpha \beta}.
\label{eq:ch2_p}
\end{equation}

In the rest frame we define the polarization four-vector as $\hat{s}^{\mu} = (0, \, \hat{\mathbf{s}})$. In a given inertial frame where the three-momentum of the particle is $\mbf{p}$, the Lorentz transformation of the polarization four-vector, $\hat{s}^{\mu}$, then becomes \cite{HiKoRy97}

\begin{equation}
s^{\mu} = \left(\dfrac{\hat{\mathbf{s}} \cdot \mathbf{p}}{m},\,\, \hat{\mathbf{s}} + \dfrac{\hat{\mathbf{s}} \cdot \mathbf{p}}{m(E + m)}\mathbf{p}\right).
\label{eq:ch2_q}
\end{equation}

Since in the limit $m \rightarrow 0$, the spin-$\frac{1}{2}$ particle is represented by the helicity state, its projection operator is redefined as 

\begin{equation}
u(\mathbf{p}, h)_{\alpha} \bar{u}(\mathbf{p}, h)_{\beta} = \left[ \dfrac{\displaystyle{\not}p}{2E}\,\dfrac{I - h \gamma_5}{2}\right]_{\alpha \beta} .
\label{eq:ch2_r}
\end{equation}

By applying the Dirac algebra the leptonic tensor becomes

\beq
L^{\text{CC}}_{\mu \nu } = \dfrac{2}{E_k E_k'}\left[k_{\mu}K_{\nu} + k_{\nu}K_{\mu} - k \cdot K g_{\mu \nu} + i\varepsilon_{\mu \nu \alpha \beta}k^{\alpha }  K^{\beta}\right]
\label{eq:ch2_s}
\eeq

where $\varepsilon_{\mu \nu \alpha \beta}$ is the antisymmetric Levi-Cevita tensor with convention $\varepsilon^{0123} = +1$, and

\beq
K^{\mu} = \dfrac{1}{2}(k'^{\mu} - h'm_l s'^{\mu});  \,\,\,\,\,\,\,\,\,\,\,\,\, {s'^{\mu} =  \dfrac{1}{m_l}(|\mathbf{k}'|, \,\, E_{k'}\hat{\mbf{k}}')},
\label{eq:ch2_t}
\eeq

where $s'^{\mu}$ is the spin polarization four-vector of the final lepton. This procedure for deriving the leptonic tensor is generally called Casimir's trick. In the zero-mass limit, we find that $K^{\mu} \rightarrow k'^{\mu}$ and hence the leptonic tensor for the NC weak transition may be deduced from Eq. \,(\ref{eq:ch2_r}): 

\beq
L^{\text{NC}}_{\mu \nu } = \dfrac{2}{E_k E_k'}\left[k_{\mu}k'_{\nu} + k_{\nu}k'_{\mu} - k \cdot k' g_{\mu \nu} + i\varepsilon_{\mu \nu \alpha \beta}k^{\alpha }  k'^{\beta}\right].\,\,\,\,\,\,
\label{eq:ch2_u}
\eeq
By employing a similar procedure to that of the leptonic case, the hadronic tensor becomes: 
\beq
W^{\mu \nu}_{\text{pol}}  =  \dfrac{1}{4} \left\lbrace \dfrac{1}{(2E_{p_{1}})(2E_{p'_{2}})}\text{Tr}\left[{J}^{\mu}(q)(\displaystyle {\not} p_1 + M_N)(I + \gamma_5 \displaystyle{\not}{s}_1)\bar{{J}^{\nu}}(q)(\displaystyle {\not} p\,'_2  + M_Y)(I + \gamma_5 \displaystyle{\not}{s\,'_2}) \right]\right\rbrace  
\label{eq:ch2_v}
\eeq
Note that in the treatment of polarized baryons, the hadronic tensor is specified as $W^{\mu \nu}_{\text{pol}} \equiv W^{\mu \nu}(q;\, p_1,\, p'_2,\, s_1,\, s'_2)$. In contrary, the unpolarized hadrons will have the tensor  $W^{\mu \nu}_{\text{unpol}} \equiv W^{\mu \nu}(q;\, p_1,\, p'_2)$. The reason for this difference lies in the fact that, in the unpolarized particle treatment, we average over the target nucleon spin state and sum over the final hyperon spin state, 
\beq
W^{\mu \nu}_{\text{unpol}}  =  \dfrac{1}{2}\left\lbrace \dfrac{1}{(2E_{p_{1}})(2E_{p'_{2}})}\text{Tr}\left[{J}^{\mu}(q)(\displaystyle {\not} p_1 + M_N)\bar{{J}^{\nu}}(q)(\displaystyle {\not} p\,'_2  + M_Y) \right]\right\rbrace ,
\label{eq:ch2_w}
\eeq
where $\bar{{J}^{\mu}}(q) = \gamma^0 {{J^{\mu}}}^{\,\dagger}(q) \gamma^0$. Eventually, by invoking energy and momentum conservation at the hadronic vertex via the Dirac delta function, the \dcs given in Eq.\,(\ref{eq:ch2_b}) may become 
\beq
\dfrac{\rm d^3 \sigma}{\rm dE_{k'}\rm d(\cos \theta ' )\rm d \Omega '_1} = \dfrac{1}{2 (2\pi)^{5}}\dfrac{G^2_F \eta^2}{2}  \chi(E_{k'})\left[\dfrac{\left( E^2_{p'_1} - M^2_K\right)^{\frac{1}{2}}}{|f'(E_{p'_1})|}\right]   L_{\mu \nu} W^{\mu \nu}.
\label{eq:ch2_x}
\eeq
Here $f(E_{p'_1})$ and $\chi(E_{k'})$ are scalar functions of $E_{p'_1}$ and $E_{k'}$, respectively, and are defined as 
\bq f(E_{p'_1}) = E_k + M - E_{k'} - E_{p'_1} - E_{p'_2}(E_{p'_1}),\,\,\,\,\,\,\, \label{eq:ch2_ga} \,\,\,\\
\chi(E_{k'}) = \left\lbrace \brr{ll}
 2\pi E^{2}_{k'} & \text{for NC process} \vspace*{0.125cm}\\
2\pi E_{k'}\left( E^2_{k'} - m^2_l\right)^{\frac{1}{2}} & \text{for CC process} \label{eq:ch2_gb}
\err\right..
\label{eq:ch2_y}\eq
Note that the weak hadronic transition current $\hat{J^{\mu}}(q)$ is not well known because of the complication that arises from the strong interaction effects at the hadronic vertex, which are not yet well understood. The next section will focus on the model-independent construction of the most general form of hadronic weak current in the context of the three-body process.

\subsection{\label{sec:level2p2}General representation of the hadronic current}
Unlike the leptonic case, the hadronic coupling vertex must be parametrized by form factors to take into account the strong interaction effects. In Ref.\,\cite{RBD64}, Bjorken and Drell decomposed the electromagnetic current operators in terms of two independent form factors. The introduction of these form factors lead to the understanding that hadrons are composite particles and the typical properties of these hadrons result from the individual contributions of the constituents. We present, for the first time, the general form of the hadronic vertex for the three-body weak transitions by extending the basic principle used by Bjorken and Drell for the two-body electromagnetic transition.
\\
\begin{figure}[h]
\centering
\includegraphics[scale=0.825]{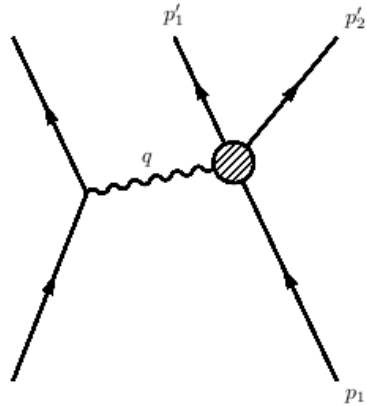} 
\caption{The Feynman diagram with a vertex of three external lines of hadrons.}
\label{fig:ch2_1}
\end{figure}

Figure\,\ref{fig:ch2_1} shows the appropriate Feynman diagram that represents the three-body process under consideration. We choose $p_1$ and $p'_2$ to be the \frma of the initial and final state baryons, respectively, and $p'_1$ to be that of the pseudoscalar meson. The hadronic vertex represents the coupling of the generalized weak hadronic current, $J^{\mu}$, with the gauge boson carrying four-momentum $q$. The weak transition current at the hadronic vertex of may be written as
\beq
\left\langle \text{K}(p'_1)\text{B}(p'_2)\left|\hat{J^{\mu}}(q) \right| \text{B}(p_1)\right\rangle \equiv \bar{u}(\mbf{p}'_2)J^{\mu}(q)u(\mbf{p}_1).
\label{eq:ch2_z}
\eeq

This model-independent derivation is done by taking advantage of the fact that the weak interaction violates the most fundamental discrete symmetries such as parity ($\mcl{P}$), time-reversal ($\mcl{T}$), charge conjugation ($\mcl{C}$), and the combination of parity and charge conjugation ($\mcl{CP}$); and hence the general expression of the current will have a relatively large number of independent parametrization form factors which possibly allow us to gain more insight into the weak interaction and the structure of hadronic weak current. In addition, unlike the electromagnetic current, the nonconservation of the weak current, resulting from the massive gauge boson exchange, does not allow the possibility of reducing the number of independent invariant amplitudes.

The derivation of the general expression of ${J^{\mu}}(q)$ is based on the global structure that the  current operator possesses. That is, ${J^{\mu}}(q)$ is a Lorentz vector and a $4 \times 4$ matrix. Note that the current operator is always sandwiched between the Dirac spinor fields. Hence we can expand ${J^{\mu}}(q)$ in terms of the bilinear covariant basis elements $\bm{\Gamma} = \{ I, \,\gamma_5, \,\gamma^{\mu}, \,\gamma_5 \gamma^{\mu}, \,\sigma^{\mu \nu} \}$, where $\sigma^{\mu \nu} = \frac{i}{2}[\gamma^{\mu},\,\gamma^{\nu}]$. Thus our first decomposition of the current operator is:
\beq
\bar{u}(\mbf{p}'_2)J^{\mu}(q)u(\mbf{p}_1)  = \bar{u}(\mbf{p}'_2)\left\lbrace \tilde{A}^{\mu}I + \tilde{B}^{\mu}\gamma_5  + \tilde{C}^{\mu \nu} \gamma_{\nu} + \tilde{D}^{\mu \nu }\gamma_5 \gamma_{\nu} + \tilde{E}^{\mu \nu \alpha}\sigma_{\nu \alpha}\right\rbrace u(\mbf{p}_1),
\label{eq:ch2_a1}
\eeq
where $\tilde{A}^{\mu},\, \tilde{B}^{\mu},\,  \tilde{C}^{\mu \nu}, \,\tilde{D}^{\mu \nu }$, and $\tilde{E}^{\mu \nu \alpha}$ are unknown tensors. However, in such transitions these unknown tensors depend on all available independent four-momenta carried by the particles participating in the interaction. Because of energy-momentum conservation at the vertex, we may only have three of them: $\{q^{\mu}, \, p_1^{\mu},\, {p'}^{\mu}_2 \}$. In addition, the metric tensor, $g^{\mu \nu}$, the Levi-Cevita tensor, $\varepsilon^{\mu \nu \alpha \beta}$, and their proper combination with the independent four-momenta are also at our disposal to further expand the unknown tensors. For instance, $\tilde{A}^{\mu}$ and $\tilde{B}^{\mu}$ are first-rank tensors and hence can be further decomposed by using the following basis elements: 
\beq
\tilde{\mcl{J}_1} = \left\lbrace q^{\mu},\,p^{\mu}_1,\,p'^{\mu}_2, \,\varepsilon^{\mu \nu \alpha \beta} q_{\nu}{p_1}_{\alpha}{p'_2}_{\beta} \right\rbrace.
\label{eq:ch2_a2}
\eeq
However, $\tilde{C}^{\mu \nu}$ and $\tilde{D}^{\mu \nu}$ are second-rank tensors and hence we use basis elements: $g^{\mu \nu},\, q^{\mu}q^{\nu}$, $\,q^{\mu}{p_1}^{\nu} \pm {p_1}^{\mu}q^{\nu}$,\, $\varepsilon^{\mu \nu \alpha \beta}q_{\alpha}{p_1}_{\beta}$, etc, with which we may construct our second set, $\tilde{\mcl{J}_2}$. The last unknown 
coefficient, $\tilde{E}^{\mu \nu \alpha}$, is the third-rank tensor. Hence by noticing that the symmetric behavior arises from the exchange of the Lorentz indices, our third set, $\tilde{\mcl{J}_3}$, may be restricted to take only basis elements of the form: $q^{\mu}(q^{\nu}{p_1}^{\alpha} - {p_1}^{\nu}q^{\alpha}),\, q^{\mu}(q^{\nu}{p'_2}^{\alpha} - {p'_2}^{\nu}q^{\alpha}),\,\,q^{\mu}({p_1}^{\nu}{p'_2}^{\alpha} - {p'_2}^{\nu}{p_1}^{\alpha}),\, q^{\mu}\varepsilon^{\nu \alpha \beta \eta}q_{\beta}{p_1}_{\eta},\, \varepsilon^{\mu \nu \alpha \beta}q_{\beta}$, etc,.

The basis elements in Eq. \,(\ref{eq:ch2_a2}) will allow us to expand $\tilde{A}^{\mu}$ as follows:
\beq
\label{eq:ch2_a3}
\bar{u}(\mbf{p}'_2)\tilde{A}^{\mu}Iu(\mbf{p}_1) \,= \,\bar{u}(\mbf{p}'_2)\left\lbrace \tilde{A}_{1}Iq^{\mu} \,\, + \,\,\tilde{A}_{2}Ip^{\mu}_1 \,\,+\,\, \tilde{A}_{3}Ip'^{\mu}_2, \,\, + \,\,\tilde{A}_{4}I\varepsilon^{\mu \nu \alpha \beta} q_{\nu}{p_1}_{\alpha}{p'_2}_{\beta}\right\rbrace u(\mbf{p}_1).
\eeq
It is quite obvious that $\bar{u}(\mbf{p}'_2)\tilde{B}^{\mu}\gamma_5u(\mbf{p}_1)$ can be expanded in a similar manner by substituting $I$ by $\gamma_5$ and $\tilde{A}_i$ by $\tilde{B}_i$. We also notice that all of the four terms in Eq.\,(\ref{eq:ch2_a3}) are independent and hence no further expansion is needed. The decomposition of $\bar{u}(\mbf{p}'_2)\tilde{C}^{\mu \nu}\gamma_{\nu}u(\mbf{p}_1)$ in terms of the second set of basis elements is
\beq
\brr{ll}
\bar{u}(\mbf{p}'_2)\tilde{C}^{\mu \nu}\gamma_{\nu}u(\mbf{p}_1) = & \bar{u}(\mbf{p}'_2)\left\lbrace \tilde{C}_1\gamma^{\mu} + \tilde{C}_2q^{\mu}\displaystyle {\not}q + \tilde{C}_3{p_1}^{\mu}{\displaystyle {\not}{p_1}} + \tilde{C}_4{p'_2}^{\mu}{\displaystyle {\not}{p\,'_2}} + \tilde{C}_5(q^{\mu}{\displaystyle {\not}{p_1}} + {p_1}^{\mu}\displaystyle {\not}q)\right.\vspace*{0.15cm}\\
&+ \tilde{C}_6(q^{\mu}{\displaystyle {\not}{p_1}} - {p_1}^{\mu}\displaystyle {\not}q) + \tilde{C}_7(q^{\mu}{\displaystyle {\not}{p\,'_2}} + {p'_2}^{\mu}\displaystyle {\not}q) + \tilde{C}_8(q^{\mu}{\displaystyle {\not}{p\,'_2}} - {p'_2}^{\mu}\displaystyle {\not}q)\vspace*{0.15cm}\\
& +\tilde{C}_9({p_1}^{\mu}{\displaystyle {\not}{p\,'_2}} + {p'_2}^{\mu}{\displaystyle {\not}{p_1}}) + \tilde{C}_{10}({p_1}^{\mu}{\displaystyle {\not}{p\,'_2}} - {p'_2}^{\mu}{\displaystyle {\not}{p_1}}) + \tilde{C}_{11}\varepsilon^{\mu \nu \alpha \beta}\gamma_{\nu}q_{\alpha}{p_1}_{\beta}\vspace*{0.15cm}\\
& + \left.\tilde{C}_{12}\varepsilon^{\mu \nu \alpha \beta}\gamma_{\nu}q_{\alpha}q_{\alpha}{p'_2}_{\beta} + \tilde{C}_{13}\varepsilon^{\mu \nu \alpha \beta}\gamma_{\nu}{p_1}_{\alpha}{p'_2}_{\beta} \right\rbrace u(\mbf{p}_1),
\err
\label{eq:ch2_a4} 
\eeq

where the Dirac slash notation $\displaystyle {\not}{a} = a^{\nu}\gamma_{\nu}$ is used. We also use the second set of basis elements to expand $\bar{u}(\mbf{p}'_2)\tilde{D}^{\mu \nu}\gamma_{5}\gamma_{\nu}u(\mbf{p}_1)$, which would have more or less similar structure. Now by exhaustively applying the Dirac algebra for on-shell particles (i.e., particles that satisfy $\displaystyle {\not}{p}u(\mbf{p}) = Mu(\mbf{p})$\,) all proportional terms can be eliminated and then the following expression may be obtained
\beq
\brr{lll}
\bar{u}(\mbf{p}'_2)\tilde{C}^{\mu \nu}\gamma_{\nu}u(\mbf{p}_1) & = & \bar{u}(\mbf{p}'_2)\left\lbrace \tilde{C}_1\gamma^{\mu} \,+\, \tilde{C}'_1 q^{\mu} \, + \, \tilde{C}'_2p^{\mu}_1 \, + \,  \tilde{C}'_3 p'^{\mu}_2 + \tilde{C}'_4\gamma_{5}\gamma^{\mu}\right.\vspace*{0.15cm}\\
& & +  \tilde{C}'_5\gamma_{5}q^{\mu} + \tilde{C}'_6\gamma_{5}{p_1}^{\mu} + \tilde{C}'_7\gamma_{5}{p'_2}^{\mu} + \tilde{C}_2q^{\mu}\displaystyle {\not}{q} \vspace*{0.15cm}\\
& &  + \left.\tilde{C}'_8{p_1}^{\mu}\displaystyle {\not}{q} + \tilde{C}'_9 {p'_2}^{\mu}\displaystyle {\not}{q} + \tilde{C}'_{10} \gamma_{5}\displaystyle {\not}{q}\gamma^{\mu}\right\rbrace u(\mbf{p}_1),
\err
\label{eq:ch2_a5}
\eeq
where $\tilde{C}_1$ and $\{\tilde{C}'_i\}_{i = 1}^{10}$ are unknown coefficients obtained by factoring out all possible independent basis elements. It is worth noting that $\tilde{C}^{\mu \nu}\gamma_{\nu}$ introduces six new expansion basis elements, $\{\gamma^{\mu}, \, \gamma_{5}\gamma^{\mu},\,q^{\mu}\displaystyle {\not}{q},\, {p_1}^{\mu}\displaystyle {\not}{q}$,\, ${p'_2}^{\mu}\displaystyle {\not}{q},\, \gamma_{5}\displaystyle {\not}{q}\gamma^{\mu}\}$, whereas the rest are proportional to the ones belonging to either $\bar{u}(\mbf{p}'_2)\tilde{A}^{\mu}Iu(\mbf{p}_1)$ or $\bar{u}(\mbf{p}'_2)\tilde{B}^{\mu}\gamma_5u(\mbf{p}_1)$. A similar procedure as for $\bar{u}(\mbf{p}'_2)\tilde{C}^{\mu \nu}\gamma_{\nu}u(\mbf{p}_1)$ also yields the following expression for $\bar{u}(\mbf{p}'_2)\tilde{D}^{\mu \nu}\gamma_5\gamma_{\nu}u(\mbf{p}_1)$:
\beq
\brr{lll}
\bar{u}(\mbf{p}'_2)\tilde{D}^{\mu \nu}\gamma_5\gamma_{\nu}u(\mbf{p}_1) & = & \bar{u}(\mbf{p}'_2)\left\lbrace \tilde{D}_1\gamma_5\gamma^{\mu} \,+\, \tilde{D}'_1 q^{\mu} \, + \, \tilde{D}'_2p^{\mu}_1 \, + \,  \tilde{D}'_3 p'^{\mu}_2 +  \tilde{D}'_4\gamma^{\mu}\right. \vspace*{0.15cm}\\
& & + \tilde{D}'_5\gamma_{5}q^{\mu} + \tilde{D}'_6\gamma_{5}{p_1}^{\mu} + \tilde{D}'_7\gamma_{5}{p'_2}^{\mu} + \tilde{D}_2q^{\mu}\gamma_{5}\displaystyle {\not}{q} +  \tilde{D}'_{8}{p_1}^{\mu}\displaystyle {\not}{q} \vspace*{0.15cm}\\
& & + \left. \tilde{D}'_{9}{p'_2}^{\mu}\displaystyle {\not}{q} + \tilde{D}'_{10}{p_1}^{\mu}\gamma_{5}\displaystyle {\not}{q} + \tilde{D}'_{11} {p'_2}^{\mu}\gamma_{5}\displaystyle {\not}{q} + \tilde{D}_{5} \displaystyle {\not}{q}\gamma^{\mu} \right\rbrace u(\mbf{p}_1).
\err
\label{eq:ch2_a6}
\eeq
As a result of the expansion of $\bar{u}(\mbf{p}'_2)\tilde{D}^{\mu \nu}\gamma_5\gamma_{\nu}u(\mbf{p}_1)$ additional contributions come through the amplitudes associated with the new basis elements: $\{q^{\mu}\gamma_{5}\displaystyle {\not}{q}$, \,${p_1}^{\mu}\gamma_{5}\displaystyle {\not}{q}$,\,  ${p'_2}^{\mu}\gamma_{5}\displaystyle {\not}{q}$,\, \,$\displaystyle {\not}{q}\gamma^{\mu}\}$. The remaining terms are proportional to the ones contained by  $\bar{u}(\mbf{p}'_2)\tilde{A}^{\mu}Iu(\mbf{p}_1)$, $\bar{u}(\mbf{p}'_2)\tilde{B}^{\mu}\gamma_5u(\mbf{p}_1)$, or $\bar{u}(\mbf{p}'_2)\tilde{C}^{\mu \nu}\gamma_{\nu}u(\mbf{p}_1)$ and hence fuse into coefficients of the same basis elements. However, the last component $\bar{u}(\mbf{p}'_2)\tilde{E}^{\mu \nu \alpha}\sigma_{\nu \alpha}u(\mbf{p}_1)$ does not carry new parameters; instead, all of them are absorbed by any of the previous ones.

Finally, the most general hadronic weak current for \spp may be expressed in terms of eighteen independent amplitudes; and in a more convenient rearrangement it may be written as:
\beq
\brr{ll}
\bar{u}(\mbf{p}'_2)J^{\mu}(q)u(\mbf{p}_1) = & \bar{u}(\mbf{p}'_2)\left\lbrace \tilde{A}^{\mu}I + \tilde{B}^{\mu}\gamma^{5} + \tilde{C}_1 \gamma^{\mu} + \tilde{C}^{\mu} \displaystyle{\not} q + \tilde{D}_1 \gamma_5 \gamma^{\mu}\right. \vspace*{0.15cm}\\
& + \left. \tilde{D}^{\mu} \gamma_5 \displaystyle {\not} q + \tilde{D}_5 \displaystyle{\not}q \gamma^{\mu} + \tilde{D}_6 \gamma_5 \displaystyle {\not} q \gamma^{\mu}\right\rbrace  u(\mbf{p}_1),
\err
\label{eq:ch2_a7}
\eeq
where
\beq
\brr{ll}
\tilde{A}^{\mu} & = \tilde{A}_1 q^{\mu} + \tilde{A}_2 p_1^{\mu} + \tilde{A}_3 {p'_2}^{\mu} + \tilde{A}_4 \varepsilon^{\mu \nu \alpha \beta}q_{\nu}{p_1}_{\alpha}{p'_2}_{\beta},\vspace*{0.15cm}\\
\tilde{B}^{\mu} & = \tilde{B}_1 q^{\mu} + \tilde{B}_2 p_1^{\mu} + \tilde{B}_3 {p'_2}^{\mu} + \tilde{B}_4 \varepsilon^{\mu \nu \alpha \beta}q_{\nu}{p_1}_{\alpha}{p'_2}_{\beta},\vspace*{0.15cm}\\
\tilde{C}^{\mu} & = \tilde{C}_2 q^{\mu} + \tilde{C}_3 p_1^{\mu} + \tilde{C}_4 {p'_2}^{\mu},\vspace*{0.15cm}\\
\tilde{D}^{\mu} & = \tilde{D}_2 q^{\mu} + \tilde{D}_3 p_1^{\mu} + \tilde{D}_4 {p'_2}^{\mu}.
\err
\label{eq:ch2_a8}
\eeq

These amplitudes are unknown. In the electromagnetic interaction, the current operator of the two-body processes can be specified by two independent form factors, which can be determined by applying a Rosenbluth separation. In contrast, in this study there are eighteen independent structure functions, which necessitate the use of a model to determine them. However, this general form of the weak current operator avoids the recalculation of the \dcs whenever we consider various reactions, introduce other models, or add more diagrams such as resonances to the Born diagram. Instead, our formalism of the \dcs can be used in more generally by only updating the eighteen invariant amplitudes of the hadronic weak current. 
\subsection{\label{sec:level2p3}Born Term Model}

What makes the development of the weak interactions of leptons relatively simple is that leptons have been identified as pointlike fermions. As a result, the standard electroweak gauge model can be used to construct the interaction Lagrangian. However, the structure of the weak interactions of hadrons is not well understood because hadrons are composite particles and hence associated strong interaction effects exist. 

In order to perform numerical calculations for the \spp processes, we introduce the Born term approximation of the hadron transition vertex. In this approximation the hadronic vertex is expanded in terms of the Born diagram with  parametrization form factors, which allows us to include the high-order contributions and circumvent the strong interaction effects. The eighteen unknown parameters in the general hadronic current given in Eq.\,(\ref{eq:ch2_a7}) will be extracted after carefully expanding the Born diagram in terms of the basis elements that are constructed in the general representation of the weak hadronic current.

\begin{table}[h]
\caption{The exchange particles and the corresponding propagators of the $s$, $t$, and $u$ channels.}
\centering
\begin{ruledtabular}
\begin{center}
% use packages: array
\begin{tabular}{lll}
Channel & Exchange particle & Propagator \\
\hline
$s$ & N = $\{\text{p}(939),\,\text{n}(939)\}$ & $\dfrac{\displaystyle {\not}{q} + \displaystyle {\not}{p}_1 + M_N}{s - M^2_N}$ \\

$t$ & K = $\{\text{K}^0(498),\, \text{K}^{+}(494)\}$ & $\dfrac{1}{\,\,\,\,\,\,t - M^2_K}$ \\
$u$ & Y = $\{\Sigma^{+}(1189),\, \Sigma^{0}(1192),\, \Sigma^{-}(1197),\,\Lambda (1116)\}$ \,\,\,\,\,\,& $\dfrac{\displaystyle {\not}{q} - \displaystyle {\not} p\,'_2 + M_Y}{u - M^2_Y}$ \\
\end{tabular}
\end{center}
\end{ruledtabular}
\label{tab:ch2_2}
\end{table}

In Refs.\,\cite{RSH75,DE80,ME76} we realize the familiarity of this model in the study of strange particle production. In general, the Born model allows three diagrams, labeled as the $s$, $t$, and $u$ channels. But at the practical level individual channels may or may not contribute to the weak hadronic current of a particular reaction. The mediators of these channels are the bound-state hadrons, and Table \ref{tab:ch2_2} gives a summary of the exchange particles of the channels along with the corresponding propagators. According to this model, the hadronic weak current operator is approximated as 

\beq
{J^{\mu}}(q) \simeq {J^{\mu}_s}(q) + {J^{\mu}_t}(q) + {J^{\mu}_u}(q),
\label{eqc6_0}
\eeq 
where ${J^{\mu}_s}(q),\,{J^{\mu}_t}(q),\,\text{and}\,{J^{\mu}_u}(q)$ are the hadronic weak current contributions from the individual channels that are physical to the process. If any of the channels allows more than one exchange particle, then we sum over the individual contributions to the channel. In some cases one of the three channels is also absent owing to conservation laws that it does not obey. Table \ref{tab:ch2_1} summarizes the above argument for the CC and strangeness conserving associated production reactions. 
\begin{table}[h]
\caption{The CC associated productions and the possible channel contributions.}
\centering
\begin{ruledtabular}
\begin{center}
% use packages: array
\begin{tabular}{cccc}
Reaction & $s$ channel & $u$ channel & $t$ channel \\ 
\hline
CC1 & $-$ & 2 & 1 \\
CC2 & 1 & 2 & $-$ \\
CC3 & 1 & 1 & 1 \\ 
CC4 & 1 & 1 & 1\\ 
\end{tabular}
\end{center}
\label{tab:ch2_1}
\end{ruledtabular}
\end{table}

In general, the Born diagram contains three type of vertices: the weak baryon coupling vertex, the weak pseudoscalar meson coupling vertex, and  the strong pseudoscalar coupling vertex. The standard representation of the weak transition current of a single baryon is: 
\begin{equation}
\left\langle \text{B}' \right| \hat{J^{\mu}}(q) \left| \text{B}\right \rangle = \overline{u}_{B'} \left\lbrace F_1 \gamma^{\mu} + \dfrac{iF_2}{2M}\sigma^{\mu \nu}q_{\nu} - G_A \gamma_5 \gamma^{\mu}\right\rbrace u_{B}. 
\end{equation}
Note that the baryon CC belong to the $\rm{SU(3)}$ octet and according to Ref. \cite{NCA63} the CVC hypothesis and the $\rm{SU(3)}$ current algebra lead to the determination of the vector form factors: $F_1$\, and\,$F_2$, of  the  weak current from the experimentally well-known electromagnetic form factors of the nucleon, ${f_i}^{p}(Q^2)$ and ${f_i}^{n}(Q^2)$. Moreover, the standard axial form factor, $G_A$, of any weak baryonic CC transition can also be determined form the nucleon axial form factor, $g_A$: 
\begin{equation}
g_A(Q^2) = \dfrac{g_A(0)}{\left[1 + 3.31 \tau (Q^2)\right]^2 },  
\end{equation}
where $g_A(0) = +1.26 \,\text{and}\, \tau (Q^2) =  {Q^2}/{4M^2}$. In Table \ref{tab:ch2_3} we present the $\rm{SU(3)}$ calculations of the vector and axial form factors of the octet baryon CC weak transitions. The determination of the form factors of the weak transitions of the pseudoscalar K-mesons is done based up on Ref. \cite{ME76}:
\bq
 \label{eqc5_42a}
\langle {\text{K}^{\lambda}(p'_1)} | \hat{J^{\mu}}(q) | {\text{K}^{\lambda'}} \rangle  = \left\lbrace \brr{ll}
 (2{p'_1}^{\mu} - q^{\mu})F_{\text{K}^{\lambda}}(q^2), & \lambda,\,\lambda' \in \{ 0, +\},\, \lambda = \lambda'. \\
(2{p'_1}^{\mu} - q^{\mu})F_{\text{K}^{\lambda, \,\lambda'}}(q^2), & \lambda, \, \lambda' \in \{ 0, +\},\,\lambda \neq \lambda'. \err \right. 
 \eq

Eventually for the strong pseudoscalar interaction vertex, $\left\langle \text{KY}|\text{N}\right\rangle =  \bar{u}_{\text{Y}}[\gamma_5 g_{\text{KYN}} ] u_{\text{N}}$, we need to find the estimated values of the strong coupling constants $g_{\text{K}\Lambda \text{N}}$ and $g_{\text{K}\Sigma \text{N}}$. Since they have not been determined from experiment, we rather adopt from Ref.\,\cite{HTH66} the $\rm{SU(3)}$ symmetry predictions of $g_{\text{K}^+\Lambda \text{p}}$ and $g_{\text{K}^+ \Sigma^0 \text{p}}$ in terms of $g_{\pi \text{NN}}$, which is experimentally well known. Then we make use of the isospin symmetry relations given in Ref.\,\cite{FXleTm01} to estimate the rest of the strong coupling constants: $g_{\text{K}^{0}\Lambda\text{n}},\,g_{\text{K}^{0}\Sigma^{0}\text{n}},\,g_{\text{K}^{0}\Sigma^{+}\text{p}},\,\text{and}\, g_{\text{K}^{+}\Sigma^{-}\text{n}}$.

\begin{table}[h]
\caption{The standard form factors for weak CC transitions of the $\rm{SU(3)}$ baryon octets.}
\centering
\begin{ruledtabular}
% use packages: array
\begin{tabular}{lccc}
Weak transition & $F_1(Q^2)$& $F_2(Q^2)$& $G_A(Q^2)$ \\
\hline
$\text{p} \rightarrow \text{n}$ & $f^{p}_1(Q^2) - f^{n}_1(Q^2)$ & $f^{p}_2(Q^2) - f^{n}_2(Q^2)$ & $g_A(Q^2)$ \\ 
$\text{p} \rightarrow \Lambda $ & $-\sqrt{\frac{3}{2}} f^{p}_1(Q^2)$ & $-\sqrt{\frac{3}{2}} f^{p}_2(Q^2)$ & $-\sqrt{\frac{1}{6}}\frac{3F + D}{F + D}g_A(Q^2)$\\ 
$\Sigma^{\pm}  \rightarrow \Lambda$ & $-\sqrt{\frac{3}{2}}f^{n}_1(Q^2)$ & $-\sqrt{\frac{3}{2}}f^{n}_2(Q^2)$ & $\sqrt{\frac{2}{3}}\frac{D}{F + D} g_A(Q^2)$ \\ 
$\Sigma^{\pm} \rightarrow \Sigma^{0}$ & $\mp\frac{1}{\sqrt{2}}[2f^{p}_1(Q^2) + f^{n}_1(Q^2)] $ & $\mp \frac{1}{\sqrt{2}}[2f^{p}_2(Q^2) + f^{n}_2(Q^2)]$ & $\mp\sqrt{2}\frac{F}{F + D}g_A(Q^2)$\\ 
\end{tabular}
\label{tab:ch2_3}
\end{ruledtabular}
\end{table}

\begin{figure}[h]
\centering
\includegraphics[scale=0.825]{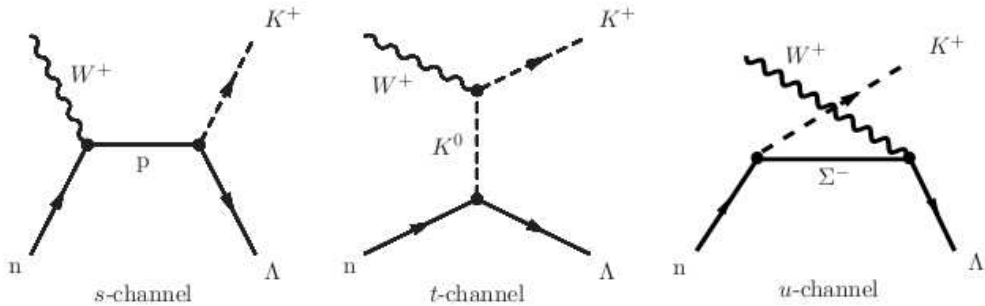} 
\caption{The Born diagram of the associated production of $K^+\Lambda$.}
\label{fig:ch2_2}
\end{figure}

For instance, we apply the Born term approximation to a typical reaction: $\nu\text{n}\,\rightarrow\,\mu^-\text{K}^+\Lambda$, which is the CC associated production. The Born diagram of this reaction is given in Fig.\,\ref{fig:ch2_2},  which shows that the three channels are allowed in the Born term approximation of CC4. Thus the weak hadronic current becomes the sum of the individual contributions of the $s$, $t$ and $u$ channels. That is 
\beq
\brr{lll}
\left\langle \text{K}^+\Lambda\left|\hat{J}^{\mu}_{\,\text{CC}}(q) \right| \text{p}\right\rangle  & \simeq & \bar{u}_{\Lambda}\left\lbrace \left[ g_{\text{K}^+\Lambda \text{p}}\gamma_5 \dfrac{\displaystyle {\not}{q} + \displaystyle {\not}{p}_1 + M_N}{s - M^2_N} \left\langle \text{p} \left|\hat{J}^{\mu}_{\text{CC}}(q)\right|\text{n}\right\rangle  \right]\right. \vspace*{0.25cm} \\
& &  \,\,\,\,\,\, + \left[ g_{\text{K}^0\Lambda \text{n}}\gamma_5 \dfrac{1}{t - M^2_{K^0}} \left\langle \text{K}^+ \left| \hat{J}^{\mu}_{\text{CC}}(q)\right|\text{K}^0 \right\rangle  \right]\vspace*{0.25cm}\\
& &  \,\,\,\,\,\, \left. + \left[ \left\langle \Lambda \left| \hat{J}^{\mu}_{\text{CC}}(q)\right| \Sigma^{-}\right\rangle \dfrac{\displaystyle {\not}{q} - \displaystyle {\not} p\,'_2 + M_{\Sigma^{-}}}{u - M^2_{\Sigma^{-}}} g_{\text{K}^+\Sigma^{-} \text{n}}\gamma_5  \right]\right\rbrace  u_{n}.
\label{eqc6_3}
\err 
\eeq

In order to extract the eighteen unknown invariant amplitudes of \equref{eq:ch2_a7} from the Born model we exhaustively expand the current contributions of the individual channels under the assumption that the fermions participating in the reaction are on-shell particles and hence we repeatedly apply the Dirac algebra until we eliminate terms which are not independent. Then we employ the method of identification in terms of the common expansion basis elements with the most general form of the weak hadronic current. Note that a similar procedure is used for other reactions, CC1, CC2, and CC3, for the calculation of the corresponding differential cross sections. 

\section{\label{sec:level3}Numerical Results}

In this section the results of numerical calculations are presented. The unknown amplitudes of the weak hadronic current are determined via the Born term approximation. In Appendix \ref{app:level2} we tabulate the extracted amplitudes for the CC reactions: CC1, CC2, CC3, and CC4. For the sake of relevance the incident energy of the neutrino is limited to the threshold energy region. Otherwise, the reliability of the Born term model becomes questionable. The general formalism is made in such a way that it allows the investigation of the angular distribution of the \dcs with respect to the outgoing kaon angle, $\theta'_1$, for the associated production reactions under consideration. %Note also that we are working in the natural units.
\begin{figure}[h]
 \centering
 \includegraphics[width = 16.50cm,bb=0 0 299 257]{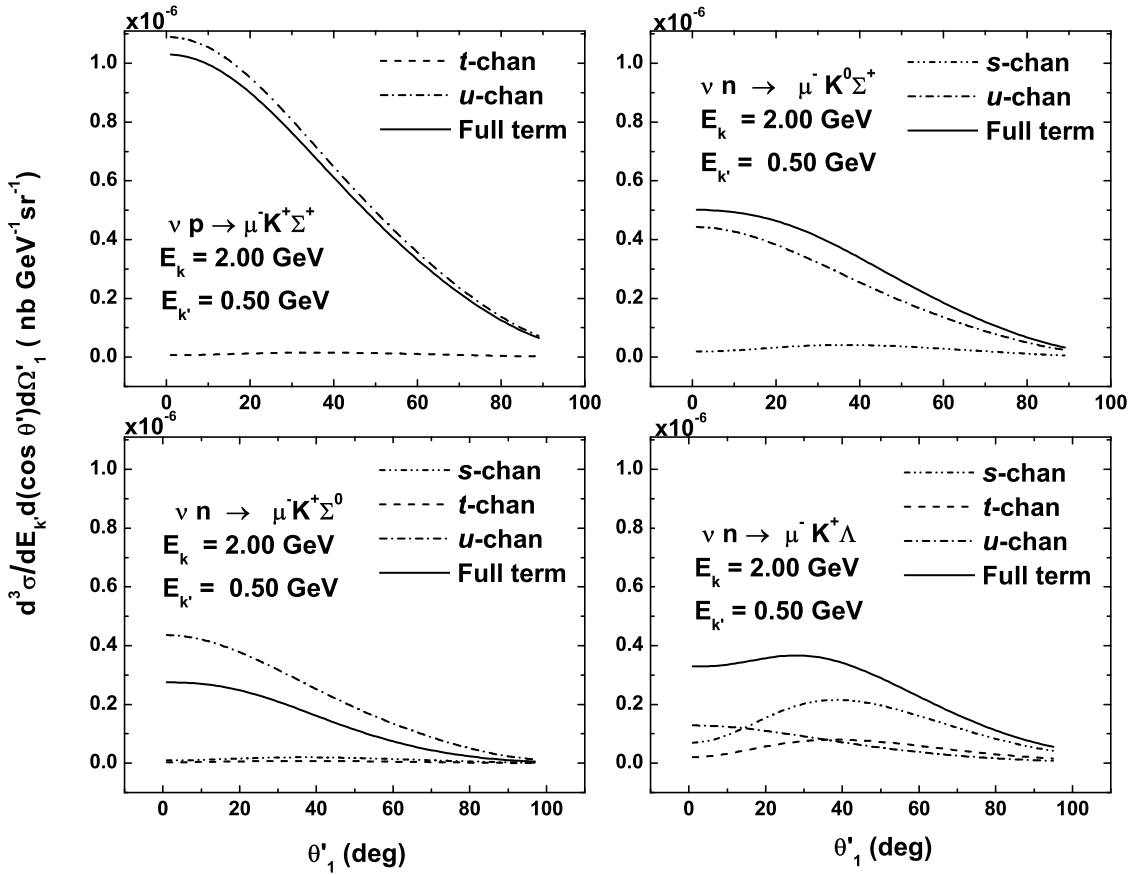}
 % all_cc1.EPS: 319x256 pixel, 72dpi, 11.25x9.03 cm, bb=0 0 319 256
\caption{The angular distribution of the differential cross sections of CC1, CC2, CC3, and CC4 with respect to the kaon angle for the first set of kinematical inputs: $E_k = 2.0\,\text{GeV}$ and $E_{k'} = 0.5\,\text{GeV}$.}
\label{fig:ch3_0}
\end{figure}

Moreover, the separate treatment of individual channels allows the comparison among their contribution to the full-term differential cross section. The contributions of $s$, $u$, and $t$ channels comes through nine, ten, and three amplitudes of the general hadronic current, respectively. Note that out of the eighteen amplitudes no contribution has come via $\tilde{A}_4, \tilde{B}_4, \tilde{C}_2,  \tilde{C}_3$, and $\tilde{C}_4$ amplitudes for the four reaction channels. 

The choice of the kinematical inputs, $E_k$ and $E_{k'}$, is made such that $Q^2$ remains constant for the entire spectrum of calculations of individual reactions. It is worth noting that the general formalism in Sec. \ref{sec:ch2p1} was established in the limit of small $Q^2$. This value of $Q^2$ is carefully selected in a region where the dominant contribution more or less comes from the lowest order Feynman diagram. The appropriate choice is $Q^2 = 0.035 \,\text{GeV}^2$. In this way two sets of input values are constructed.

\subsection{\label{sec:level3a}Differential Cross Section for Unpolarized Baryons}

We first investigate the case in which the target nucleon and the final state hyperon are considered to be unpolarized. Hence we average over and sum over the spin states of the nucleon and hyperon, respectively. This condition only affects the hadronic tensor by leaving the rest unchanged. The numerical results of the CC reactions are presented for the two sets of kinematical inputs by fixing the value of $Q^2$ at $0.035\,\text{GeV}^2$. 

\begin{figure}[h]
 \centering
 \includegraphics[width = 16.50cm,bb=0 0 299 257]{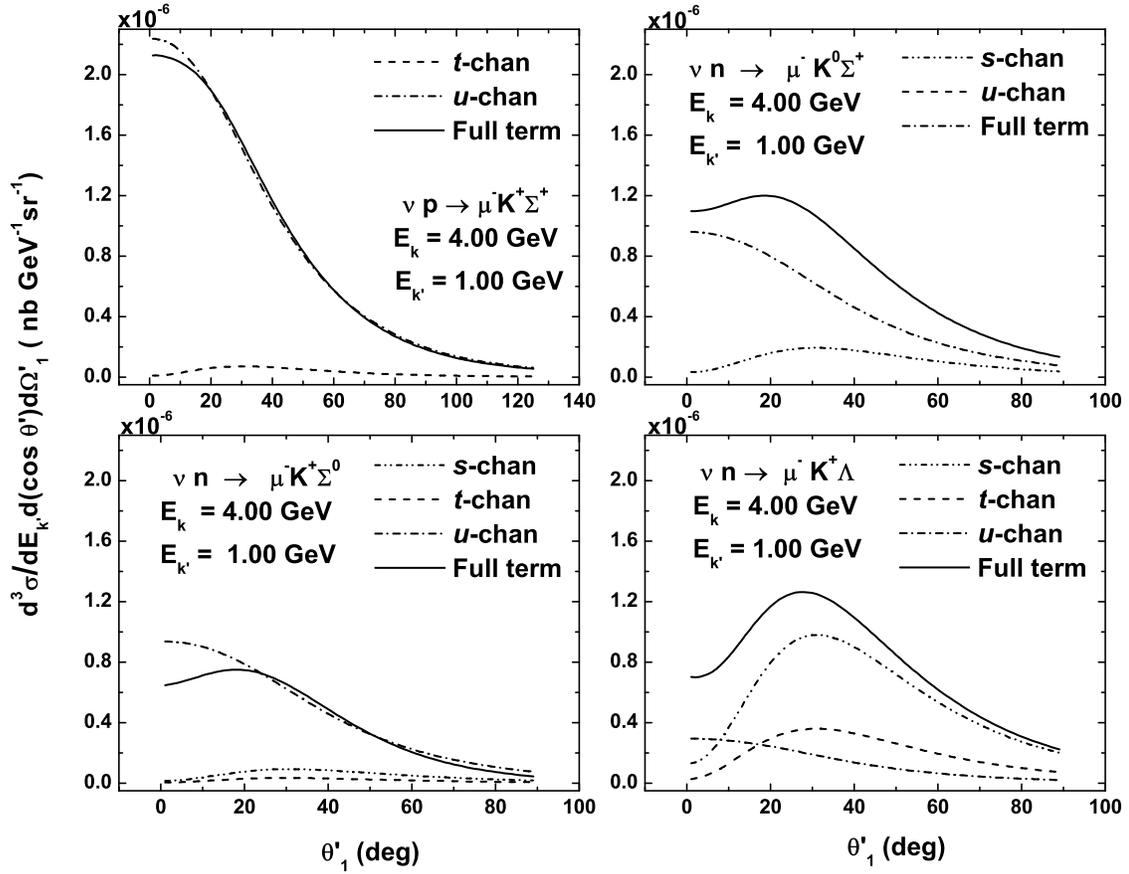}
 % all_cc2.EPS: 309x247 pixel, 72dpi, 10.90x8.71 cm, bb=0 0 309 247
\caption{The angular distribution of the differential cross sections of CC1, CC2, CC3, and CC4 with respect to the kaon angle for the second set of kinematical inputs: $E_k = 4.0\,\text{GeV}$ and $E_{k'} = 1.0\,\text{GeV}$.}
\label{fig:ch3_1}
\end{figure}

Figures \ref{fig:ch3_0} and \ref{fig:ch3_1} contain the plots of the differential cross sections of the four CC reactions for the unpolarized baryons. Each plot shows the curves of individual contributions of the channels as well as the full-term cross sections. The $u$ channel clearly dominates the other channels for CC1, CC2, and CC3 and hence the curve of the full-term cross section is forward peaked. However, for all processes the contribution of the $t$ channel is negligibly small in comparison to the others. In the CC4, however, the $s$ channel becomes significantly the dominant channel, which in turn shifts the peak of the full-term curve away from the forward angle.

\begin{figure}[h]
 \centering
 \includegraphics[width = 16.50cm,bb=0 0 319 256]{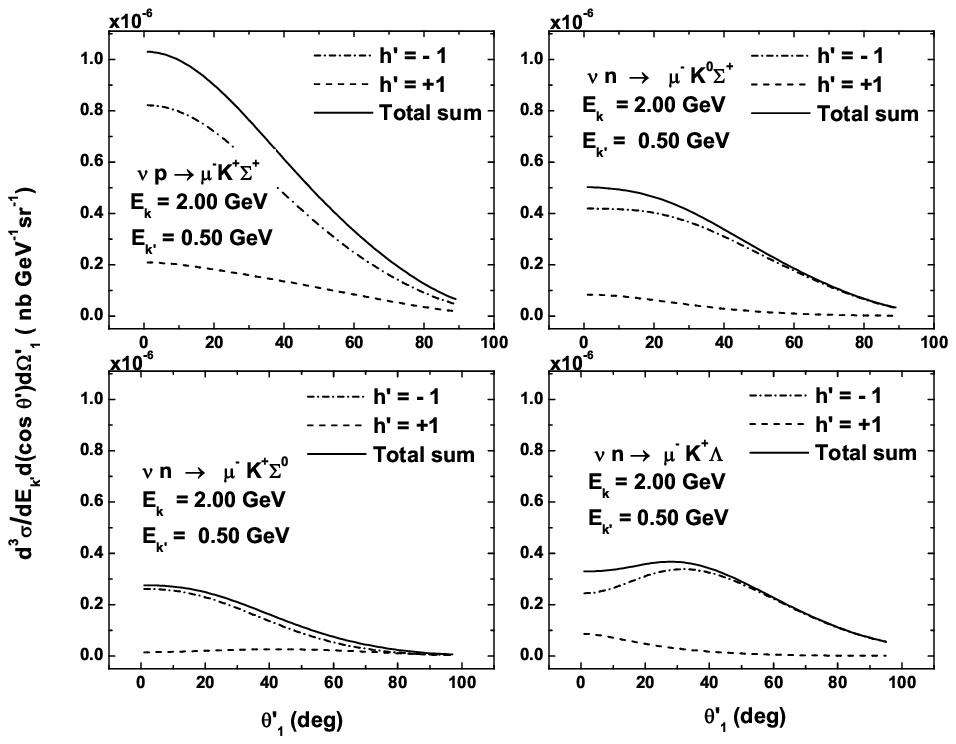}
 % all_cc_hlc.EPS: 319x256 pixel, 72dpi, 11.25x9.03 cm, bb=0 0 319 256
\caption{Comparison of the contributions of the helicity states of the outgoing muon to the differential cross sections of CC1, CC2, CC3, and CC4 for the first set of kinematical inputs.}
\label{fig:ch3_2}
\end{figure}

\begin{figure}[h]
 \centering
 \includegraphics[width = 16.50cm,bb=0 0 319 256]{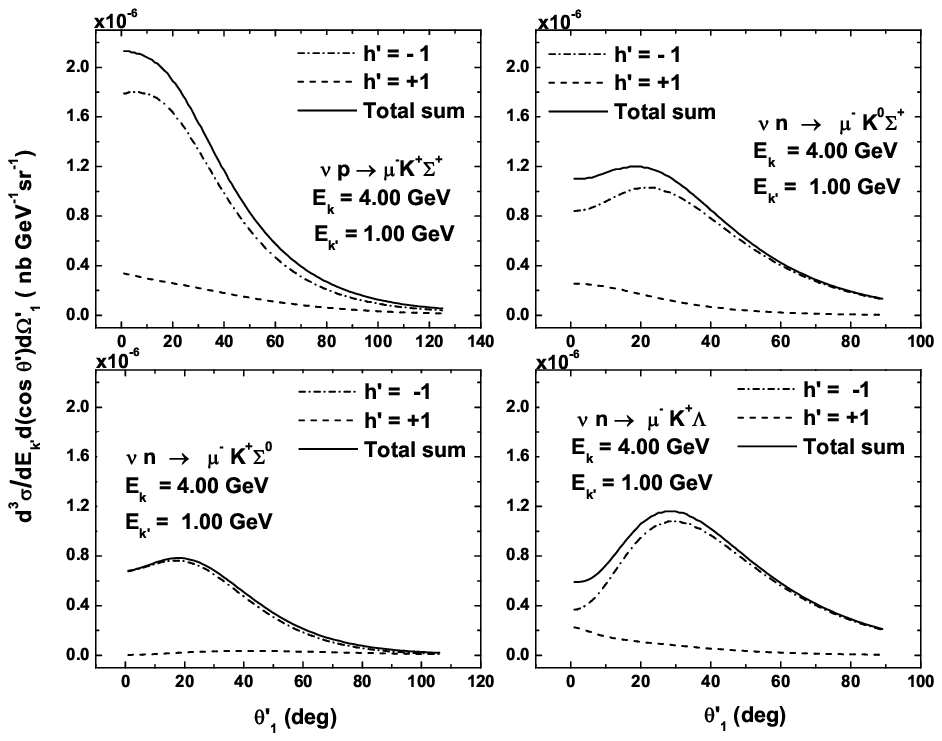}
 % all_cc_hlc.EPS: 319x256 pixel, 72dpi, 11.25x9.03 cm, bb=0 0 319 256\
\caption{Comparison of the contributions of the helicity states of the outgoing muon to the differential cross sections of CC1, CC2, CC3, and CC4 for the second set of kinematical inputs.}
\label{fig:ch3_3}
\end{figure}

For unpolarized baryons the contributions of the helicity states of the outgoing muon are compared in Figs.\,\ref{fig:ch3_2} and \ref{fig:ch3_3}. Here again we treat the participating baryons as unpolarized particles. Both figures clearly show the dominance of the contribution of the negative-helicity state over the positive one. Especially at the peaks of the curves the dominance is more significant, and hence almost the total contribution to the \dcs comes from the negative-helicity state of the muon. If we compare the muon mass to its energy we clearly see that the left-handedness of leptons is a high-energy phenomenon. Note that the positive helicity survives partly owing to the nonzero mass of the muon.

\subsection{\label{sec:level3b}Differential Cross Section for Polarized Baryons}

For polarized baryons we can investigate the sensitivity of the differential cross section to the various spin polarizations of the baryons. For convenience, by referring to Fig.\,\ref{fig:app1_0} in Appendix\,\ref{app:app1}, we choose the Cartesian coordinates of the leptonic plane to specify the directions of the three-vectors, $\hat{\bm{s}}_1$ and $\hat{\bm{s}}'_2$, of the rest-frame spin polarizations of the target nucleon and the outgoing hyperon, respectively. The three possible polarization axes are $\hat{x}$, $\hat{y}$, and $\hat{z}$ and hence there are nine possible combinations in choosing these axes for the nucleon and hyperon. As a consequence we investigate nine spin observables:

\beq
\textbf{D} = % use packages: array
\left( \begin{array}{lll}
\textbf{D}_{\text{xx}} & \textbf{D}_{\text{xy}} & \textbf{D}_{\text{xz}} \\ 
\textbf{D}_{\text{yx}} & \textbf{D}_{\text{yy}} & \textbf{D}_{\text{yz}} \\ 
\textbf{D}_{\text{zx}} & \textbf{D}_{\text{zy}} & \textbf{D}_{\text{zz}}
\end{array}\right),
\label{eq:ch3_0}
\eeq
where, for instance, the spin observable $\textbf{D}_{\text{xy}}$ is defined as
\beq
\textbf{D}_{\text{xy}} = \dfrac{\rm d^3 \sigma (\hat{\bm{s}}_1 = \hat{x};\,\,\hat{\bm{s}}'_2 = \hat{y})}{\rm dE_{k'}\rm d(\cos \theta ' )\rm d \Omega '_1}.
\label{eq:ch3_1}
\eeq 

\begin{figure}[h]
 \centering
 \includegraphics[width = 16.50cm,bb=0 0 299 257]{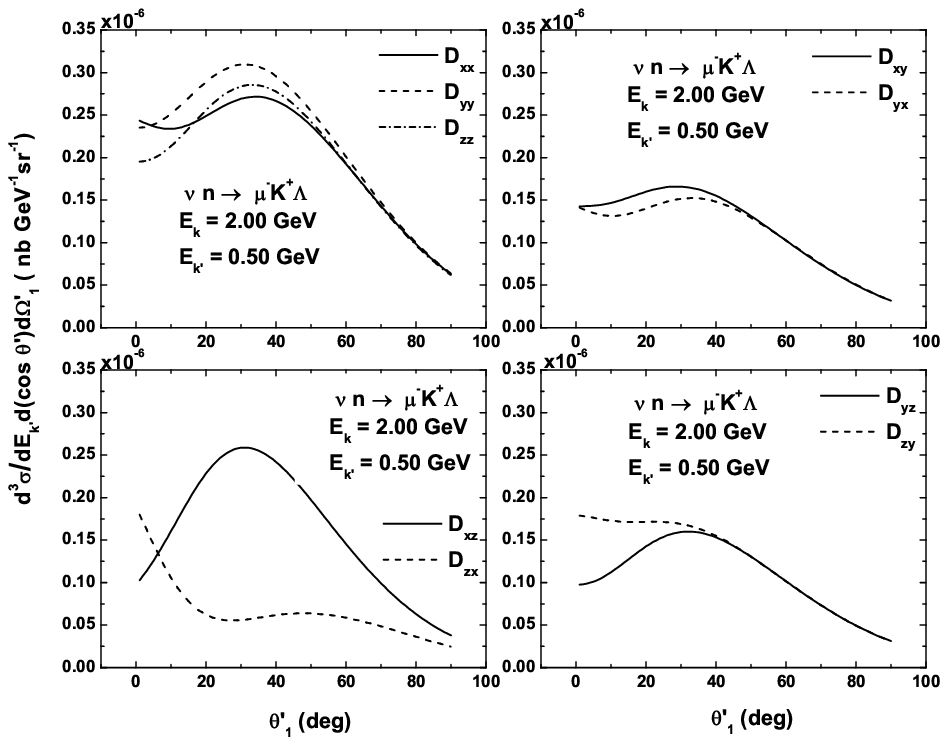}
 % cc4_D_xyz.EPS: 296x246 pixel, 72dpi, 10.44x8.68 cm, bb=0 0 296 246
\caption{The full term differential cross sections of the CC4 for the case of polarized baryons.}
\label{fig:ch3_4}
\end{figure}

\begin{figure}[h]
 \centering
 \includegraphics[width = 16.50cm,bb=0 0 299 257]{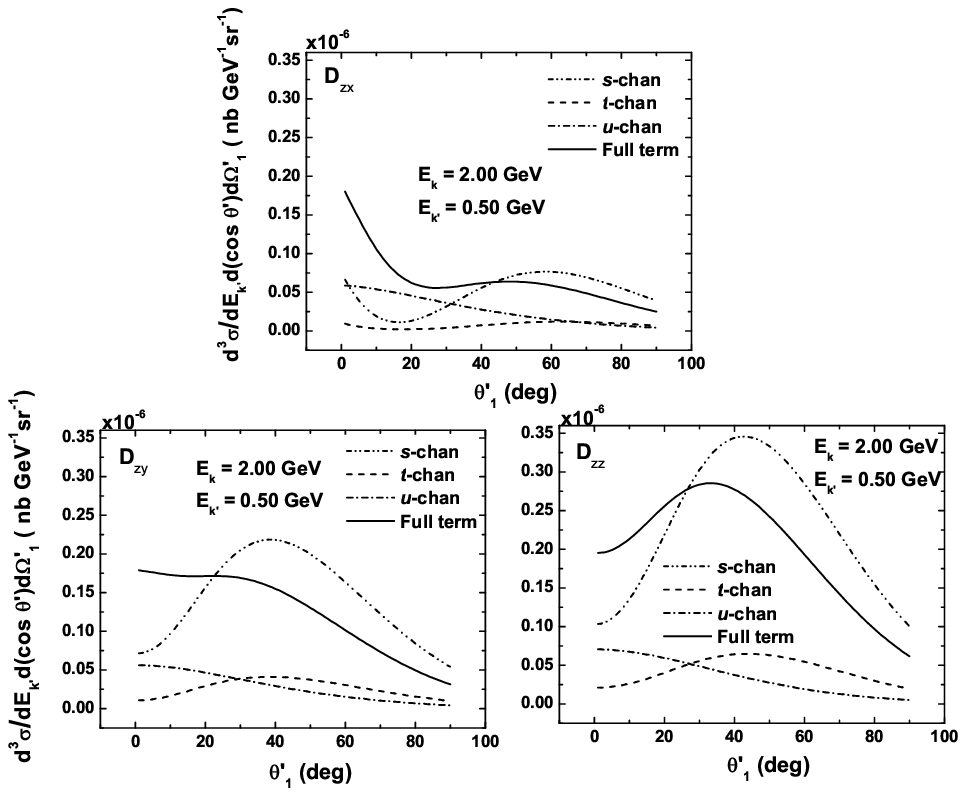}
 % stu_III_zx_zy_zz.EPS: 299x257 pixel, 72dpi, 10.55x9.07 cm, bb=0 0 299 257
\caption{The contributions of the channels of the Born diagram to the spin observables: $\textbf{D}_{\text{zx}}$, $\textbf{D}_{\text{zy}}$, and $\textbf{D}_{\text{zz}}$ for the reaction CC4.}
\label{fig:ch3_5}
\end{figure}

Since it is experimentally feasible to measure the polarization of the $\Lambda$-hyperon \cite{RSH75}, we present the plots for the CC4 for the polarized nucleon and hyperon. Figure \ref{fig:ch3_4} shows all the curves of the nine spin observables of the CC4 for the first set of input kinematics. Figure \ref{fig:ch3_5} presents the plots that give information about the contribution of the channels of the Born diagram for specific spin observables. 


\section{Discussion}

Without the knowledge of the $\rm{SU(3)}$ symmetry we can determine the invariant amplitudes of the hadronic current that comes through the $s$ channel by using the $\rm{SU(2)}$ isospin symmetry and hence this channel may not be considered as the best candidate to test $\rm{SU(3)}$. So we can safely state that the dynamics of $\rm{SU(3)}$ symmetry strongly comes through $u$ and $t$ channels. Thus CC1, CC2, and CC3 reactions are sensitive to $\rm{SU(3)}$ dynamics.

In summary, one of the peculiar features of these plots is that almost all of the differential cross sections attain their peaks at forward angles. On the other hand, the curves display the rapid decrement as the kaon angle increases. It is also clearly shown that the $u$ channel dominates all the other channels in CC1, CC2, and CC3. However, for CC4 the contribution of the $s$ channel dominantly influence the shape of the curve of the full term cross section. 

In general, the cross section increases with $E_k$ in all four reactions. The presence of the $s$ channel results in the shifting of the peak of the \dcs at forward angle toward the right. Apart from that, all of the curves have cutoff points at higher muon angle, which arises from the nonzero mass of the muon. Moreover, the numerical results show that, at forward angle, the angular distribution of the \dcs for CC1 is the highest of all, whereas CC3 gives the lowest cross section and CC2 and CC4 fall between the two.

The figures for the polarized baryons display the sensitivity of the cross section curves to the choice of the polarization axes. For instance, the differential cross sections become relatively large when the same polarization axis is picked for both particles. In addition, the plots in Fig. \ref{fig:ch3_5} indicate that the matrix in Eq. (\ref{eq:ch3_0}) is not symmetric for the full term cross section owing to the presence of the $s$ channel. However, the $t$ channel alone gives the symmetric matrix, whereas for the $u$ channel it gets somewhat close to symmetric. 

\subsection{Comparison with previous calculations}

To validate the general Lorentz-invariant representation of the hadronic current operator presented here, it is necessary to compare our results with similar calculations presented in Refs. \cite{RSH75,DE80}. To this end we calculate
\begin{equation}
\dfrac{\rm{d}\sigma (E_k,W)}{\rm{d} W} = \int_0^{Q^2_{\text{max}}}\rm{d} Q^2 \int_0^{2\pi}\rm{d}\phi ' _1 \int_0^{\pi}\rm{d}\theta '_1 \sin \theta ' _1  \dfrac{\rm{d}^4 \sigma}{ \rm{d}W\, \rm{d}Q^2 \,\rm{d}(cos\theta '_1)\, \rm{d}\phi ' _1}
\label{eq:ch4_1}
\end{equation}
where the integrand is given by Eq. (28) in Ref. \cite{RSH75} and $Q^2_{\text{max}}$ is given in Ref. \cite{DE80}. This numerical calculation has been done using a 40-point Gaussian integration, which has shown good convergence. In this expression we employed our newly derived form of the hadronic current operator, together with the Born term amplitudes and the coupling constants as given in Ref. \cite{RSH75}. 

\begin{figure}[h]
\includegraphics[scale=1,bb=0 0 299 257]{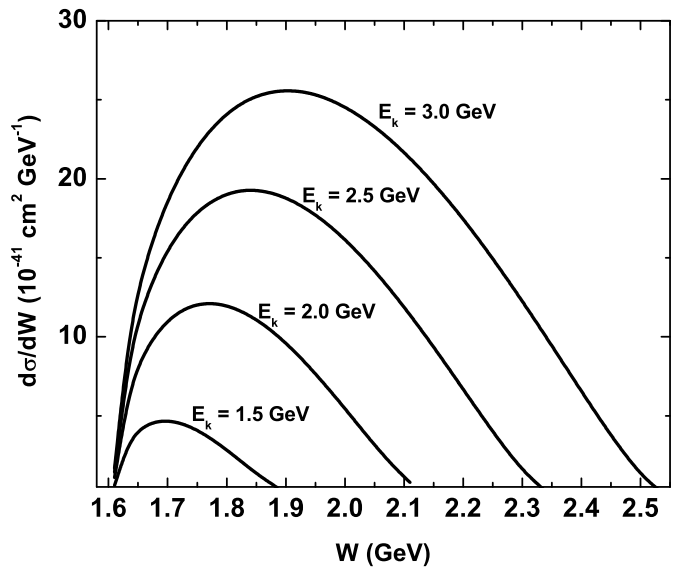}
\caption{The distribution of the differential cross section, $\rm{d}\sigma (E_k, W) / \rm{d}W$, with respect to the hadronic center-of-mass energy, $W$, for the CC reaction $\nu n \rightarrow \mu^- K^+ \Lambda$, at the neutrino incident energies, $E_k = 1.5,\,\, 2.0,\,\, 2.5$,\,and\, $3.0$ GeV. }
\label{fig:ch3_6}
\end{figure}

The results are presented in Fig. \ref{fig:ch3_6}. We can see that our results are virtually identical to those given in Fig. 5 of Ref. \cite{RSH75}. Not only do we reproduce the shape of the cross sections, we also obtain peak values practically identical to those in \cite{RSH75}. This gives us confidence that our general Lorentz-invariant form of the hadronic current operator is indeed correct. Note that a meaningful comparison with data is not really possible since we (and also the authors of Refs. \cite{RSH75,DE80}) restricted our calculations to Born-term contributions only.

\section{\label{sec:level4}Conclusions}

We have developed the most general relativistic formalism of the differential cross section for the neutrino-induced weak production of strange particles. The derivation is made for CC reactions to investigate the angular distribution of the \dcs with respect to the outgoing kaon angle. The leptonic tensor can be calculated in the framework of electroweak theory. The helicity dependence comes through the antisymmetric component of the leptonic tensor. The hadronic part of the reaction is influenced by the strong interaction, for which a perturbative calculation is not possible. Therefore, we have developed a new model-independent approach to calculate the invariant matrix element by deriving the most general weak hadronic current in terms of eighteen parametrization form factors. 

We have shown, by using the typical CC reaction, that it is possible to virtually reproduce results presented in Ref. \cite{RSH75} with the inclusion of the newly derived  weak hadronic current. At this stage it is not possible to directly compare our results with data since we have not included effects such as resonances and nuclear effects. 

The results presented in Sec. \ref{sec:level3} are entirely based on the Born term contribution. The main reason was to show that it is possible to extract the invariant amplitudes in the general expansion, when a specific model for the interaction is adopted. The authors of Refs. \cite{SngVac} and \cite{AMER77} also presented detailed models for neutrino-induced reactions. Specifically, they showed that resonances, background and nuclear effects are important for a description of the data. Note, however, that the authors of Ref. \cite{SngVac} did not consider associated production as we do here. In our analysis there is a slight shift in emphasis. We do not concentrate specifically on trying to describe the data, but rather on developing a general model for the hadronic tensor using general arguments based on Lorentz invariance. This is similar in spirit to the method developed for the description of the meson photoproduction in Ref. \cite{CGLN57}. However, owing to the large number of invariant amplitudes it is necessary for us to resort to a model to determine them. For this purpose we used a simple model including the Born terms.  

Future work and quantitative comparison with experiment will demand that we include all other effects as clearly pointed out in the references mentioned here. However, the true merit of the analysis presented here is that it is based on general principles and that we do not need to rederive all cross section formulas. Any new model additions will only mean that we have to update the relevant subroutines in our codes. This analysis also provides another example of how general principles can be used to derive the general form of a fundamental quantity such as the hadronic tensor. This approach is widely used in photo- and electroproduction and we have have now shown that is also possible for neutrino-induced associated production.

Finally, our general formalism has shown that the \dcs is sensitive to the spin polarizations of the nucleon and hyperon. In the future paper we will include the resonance and background contributions that are relevant in the threshold energy region. The CC \spp reactions having $u$ channel dominance can be used to test the validity of $\rm{SU(3)}$ symmetry. Therefore, we hope that, our predictions motivate the undertaking of further experiments in the area of neutrino and hadron physics.

\begin{acknowledgments}

This work is financially supported in part by the African Institute for Mathematical Sciences, by the University of Stellenbosch, and by the National Research Foundation of South Africa. B.I.S.v.d.V gratefully acknowledges the financial support of the National Research Foundation of South Africa under Grant No. GUN 2048567.

\end{acknowledgments} 

\appendix

\section{\label{app:app1}  KINEMATICS}

\begin{figure}[h]
\centering
\includegraphics[scale =0.825]{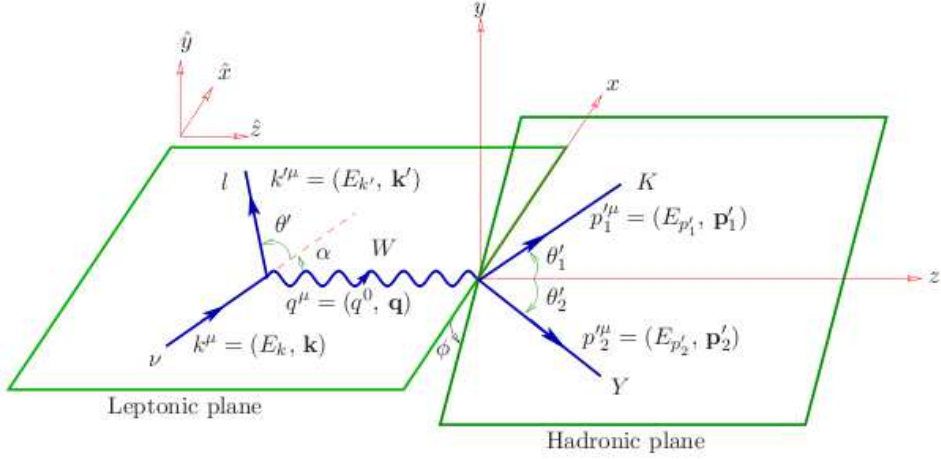}
 % lep_had_planes2.eps: 1179666x1179666 pixel, 300dpi, 9987.84x9987.84 cm, bb=0 0 628 426
\caption{kinematics of the neutrino-induced weak production reactions, $\nu N \rightarrow l K^{} Y$, in the rest frame of the target nucleon. }
\label{fig:app1_0}
\end{figure}

The kinematics of the reaction is developed in terms of the relativistic framework in the laboratory frame of reference. It is more convenient to describe the kinematics in two planes: a leptonic (scattering) plane and a hadronic (production) plane as shown in Fig.\,\ref{fig:app1_0}. The leptonic plane contains the incoming neutrino and the outgoing lepton. The gauge boson, that mediates the interaction, has the three-vector momentum transfer directed along the $z$ axis. 

The hadronic plane is defined in such a way that it is the $\rm {SO(3)}$ transformation of the leptonic plane about the $z$ axis by an angle $\phi$ and it contains the hadronic final state particles produced by the neutrino incident on the target nucleon. It is worth noting that the kinematics of the \dcs depends on the masses, energies, and momenta of the particles involved in the interaction. This dependence would be shown in the general expression of the phase space factor. 

In the leptonic plane, $\mbf{k}$ and $\mbf{k}'$ become
\beq
\brr{ll}
\mbf{k} & = E_k \left( \sin \alpha,\,\,0,\,\,\cos \alpha \right) \vspace*{0.15cm}\\
\mbf{k}' & = (E^2_{k'} - m_{\mu})^{\frac{1}{2}}\left( \sin (\alpha+\theta '),\,\,0,\,\,\cos (\alpha+\theta')\right) .
\err
\label{eqc3_5}
\eeq
Since the three-vector momentum transfer $\mathbf{q} = \mathbf{k} - \mathbf{k}'$, which is carried by the gauge boson, is defined in such a away that $\mathbf{q} = (0,\, 0,\, q_z)$, this in turn allows the determination of the incident angle $\alpha$ from input variables $E_k, \,E_{k'},\,\theta'$:
\beq
\sin \alpha = \pm |A|/(A^2 + B^2)^{\frac{1}{2}},\,\,\,\,\,\,\,\, \cos \alpha = \pm |B|/({A^2 + B^2})^{\frac{1}{2}}
\label{eqc3_6}
\eeq
where 
\beq
A = E_k - (E^2_{k'} - m_{\mu})^{\frac{1}{2}} \cos \theta ',\,\,\,\, B = (E^2_{k'} - m_{\mu})^{\frac{1}{2}} \sin \theta '.
\label{eqc3_7}
\eeq

In the hadronic plane, there are two outgoing particles with three-momenta $\mbf{p}'_1$ and $\mbf{p}'_2$ at angles $\theta'_1$ and $\theta'_2$ from the $z$ axis. Since this plane may be connected to the leptonic plane via the $\rm{SO(3)}$ group transformation, we can obtain the following expressions of $\mbf{p}'_1$ and $\mbf{p}'_2$ 
\beq
\brr{ll}
\mbf{p}'_1 &= (E^2_{p'_1} - M^2_{K})^{\frac{1}{2}}\left( \sin \theta'_1 \cos \phi,\,\,\sin \theta'_1 \sin \phi,\,\,\cos \theta'_1\right) \vspace*{0.15cm}\\
\mbf{p}'_2 &= (E^2_{p'_2} - M^2_{Y})^{\frac{1}{2}}\left( -\sin \theta'_2 \cos \phi,\,\,-\sin \theta'_2 \sin \phi,\,\,\cos \theta'_2 \right).
\err
\label{eqc3_9}
\eeq
In momentum space, the following spherical coordinate treatments can be applied for the final state particles $l(k')$, $K(p'_1)$, and $Y(p'_2)$:
\beq
\brr{ll}
\rm d^3 \mbf{k}' & = E_{k'}(E^2_{k'} - m^2_{\mu})^{\frac{1}{2}}\rm dE_{k'}d\Omega(\phi', \theta ') \vspace*{0.15cm}\\
\rm d^3 \mbf{p}'_1 & = E_{p'_1}(E^2_{p'_1} - M^2_{K})^{\frac{1}{2}} \rm dE_{p'_1}d\Omega(\phi'_1,\theta '_1) \vspace*{0.15cm}\\
\rm d^3 \mbf{p}'_2 & = E_{p'_2}(E^2_{p'_2} - M^2_{Y})^{\frac{1}{2}}\rm dE_{p'_2}d\Omega(\phi'_2,\theta '_2),
\err
\label{eqc3_10}
\eeq
where $\rm d\Omega(\phi, \theta) =  d(\cos \theta) d\phi$. Note that, we have chosen
the values of $\phi'_1$ and $\phi'_2$ such that $\phi'_1 = \phi'_2 - \phi = \phi$. The most important problem we encounter in the derivation of the cross section is the one with the four-dimensional Dirac delta function: $(2\pi)^4 \delta(k - k' + p_1 - p'_1 - p'_2)$. Based on the properties associated with the delta function, the space component enforces the conservation of three-momentum at the hadronic vertex of the Feynman diagram of the reaction. As a result, integrating over the Dirac delta function of Eq.\,(\ref{eq:ch2_b}) in Sec. \ref{sec:ch2p1} with respect to $\mbf{p}'_2$ gives rise to the following conservation relation at the hadronic vertex:
\beq
 \mbf{p}'_2 = \mbf{q} + \mbf{p} - \mbf{p}'_1.
\label{eqc3_12}
\eeq
Then the delta function becomes a function of two variables$-$ $E_{p'_1}$ and $E_{p'_2}$, which are not independent. However, owing to the relativistic energy-momentum relation we can use Eq.\,(\ref{eqc3_12}) to express $E_{p'_2}$ in terms of  $E_{p'_1}$. Thus the Dirac delta function becomes $
[f(E_{p'_1})]$, where $f(E_{p'_1}) = E_k - E_{k'} + E_p - E_{p'_1} - E_{p'_2}$. Therefore, after a few algebraic steps $f(E_{p'_1})$ becomes
\beq
f(z) = a - z - \left(b + z^2 - c \left[ z^2 - {M_K}^2 \right]^{\frac{1}{2}} \right)^{\frac{1}{2}},
\label{eqc3_13}
\eeq
where
\beq
z = E_{p'_1},\,\,\, a = E_k - E_{k}' + M,\,\,\, b = |\mbf{q}|^2 - {M_K}^2 + {M_Y}^2,\,\,\, c = 2 |\mbf{q}| \cos \theta'_1.
\label{eqc3_14}
\eeq 
Once again, by invoking the property of the Dirac delta function, it is possible to demand the conservation of energy at the hadronic vertex, which allows us the determine $E_{p'_1}$ in terms of the kinematical inputs of the reaction: $\{E_k,\, E_{k'}, \,\theta',\,\theta'_1,\, \phi\}$. For this paper we set $\theta'$ and $\phi$ to $0.5$ and $0.0$ deg, respectively. Therefore, this specification allows us to investigate the angular distribution of the differential cross section as a function of the outgoing kaon angle, $\theta'_1$. We also define the most appropriate kinematical quantities, which are called the Lorentz invariant Mandelstam variables:
\bq
\label{eqc3_17}
s = (q + p_1)^2,\,\,\,\,\, t = (q - p'_1)^2,\,\,\,\,\, u = (q - p'_2)^2.
\eq

\section{\label{app:level2} EXTRACTION OF THE INVARIANT AMPLITUDES}

Here we present the Born model extracted values of the structure functions of the general representation of the weak hadronic current for the neutrino-induced CC associated productions: CC1, CC2, CC3, and CC4 in Tables \ref{tab:app3_0}, \ref{tab:app3_1}, \ref{tab:app3_2}, and \ref{tab:app3_3}, respectively.  
\begin{table}[h]
\caption{Extraction of the unknown amplitudes for the CC1 reaction process.}
\centering
 \begin{footnotesize}
\begin{ruledtabular}
\begin{center}
% use packages: array
\begin{tabular}{cccc}
Amplitudes & $u$ channel ($\Lambda$) & $u$ channel ($\Sigma^0$)  & $t$ channel ($\text{K}^0$) \\
\hline
$\tilde{A}_1$ & $2{G^{(\Lambda)}_u}^{\footnote{\begin{scriptsize}$G_t = g_{\text{K}^0 \Sigma^+ \text{p}} (t - M^2_{K^0})^{-1},\,\,\,G^{(\Lambda)}_u  = g_{\text{K}^+ \Lambda \text{p}} (u - M^2_{\Lambda})^{-1},\,\,\,G^{(\Sigma^0)}_u  = g_{\text{K}^+ \Sigma^0 \text{p}} (u - M^2_{\Sigma^0})^{-1}$. \end{scriptsize} }} {G^{\begin{scriptsize}Y_1\end{scriptsize}}_A}^{\footnote{\begin{scriptsize}$F^{Y_1}_i(Q^2) = -\sqrt{\frac{3}{2}}f^{n}_i(Q^2),\,G^{Y_1}_A(Q^2) = \sqrt{\frac{2}{3}}\frac{D}{F + D} g_A(Q^2),\,F^{Y_2}_i(Q^2) = -\frac{1}{\sqrt{2}}(2f^{p}_i(Q^2) + f^{n}_i(Q^2)),\\ G^{Y_2}_A(Q^2) = -\sqrt{2}\frac{F}{F + D}g_A(Q^2)
$.  \end{scriptsize}}}$ & $2{G^{(\Sigma^0)}_u}^{{}^a}{G^{\begin{scriptsize}Y_2\end{scriptsize}}_A}^{{}^b}$ & $-$ \\
\hline
$\tilde{A}_2$ & $-$ & $-$ & $-$ \\
\hline
$\tilde{A}_3$ & $-2G^{(\Lambda)}_uG^{\begin{scriptsize}Y_1\end{scriptsize}}_A$ & $- 2G^{(\Sigma^0)}_uG^{\begin{scriptsize}Y_2\end{scriptsize}}_A$ & $-$\\
\hline
$\tilde{A}_4$ & $-$ & $-$ & $-$\\ 
\hline
$\tilde{B}_1$ & $G^{(\Lambda)}_u\left[2{F^{\begin{scriptsize}Y_1\end{scriptsize}}_1}^{{}^b} + {\bm{\Delta}M_2}^{\footnote{\begin{scriptsize}$\bm{\Delta}M_1 = M_{\Sigma^+} - M_{\Sigma^0},\,\,\bm{\Delta}M_2 = M_{\Sigma^+} - M_{\Lambda},\,\,\bm{\Sigma}_{M_2} = M_{\Sigma^+} + M_{\Sigma^0},\,\,\bm{\Sigma}_{M_3} = M_{\Sigma^+} + M_{\Lambda}$.\end{scriptsize} }} \dfrac{{F^{\begin{scriptsize}Y_1\end{scriptsize}}_2}^{{}^b}}{2M}\right] $ & $G^{(\Sigma^0)}_u\left[ 2{F^{\begin{scriptsize}Y_2\end{scriptsize}}_1}^{{}^b} + {\bm{\Delta}M_1}^{{}^c}\dfrac{{F^{\begin{scriptsize}Y_2\end{scriptsize}}_2}^{{}^b}}{2M}  \right]$ & ${G_t}^{{}^a}F_{K^{+,\,0}}$ \\ 
\hline
$\tilde{B}_2$ & $-$ & $-$ & $2{G_t}F_{K^{+,\,0}}$\\
\hline  
$\tilde{B}_3$ & $-2G^{(\Lambda)}_uF^{\begin{scriptsize}Y_1\end{scriptsize}}_1$ & $-2G^{(\Sigma^0)}_uF^{\begin{scriptsize}Y_2\end{scriptsize}}_1$ & $-2{G_t}F_{K^{+,\,0}}$ \\
\hline
$\tilde{B}_4$ & $-$ & $-$ & $-$\\ 
\hline
$\tilde{C}_1$ & $G^{(\Lambda)}_u{\bm{\Delta}M_2}G^{\begin{scriptsize}Y_1\end{scriptsize}}_A$ & $G^{(\Sigma^0)}_u {\bm{\Delta}M_1}G^{\begin{scriptsize}Y_2\end{scriptsize}}_A$ &$-$ \\
\hline
$\tilde{C}_2$ & $-$ & $-$ & $-$ \\ 
\hline
$\tilde{C}_3$ & $-$ & $-$ & $-$ \\ 
\hline
$\tilde{C}_4$ & $-$ & $-$ & $-$ \\ 
\hline
$\tilde{D}_1$ & $G^{(\Lambda)}_u\left[ {q'^2}^{\footnote{\begin{scriptsize}$q'^2 = ({q^2}/{2} - q \cdot p'_2)$.            \end{scriptsize} }}  \dfrac{F^{\begin{scriptsize}Y_1\end{scriptsize}}_2}{M} - {\bm{\Sigma}_{M_3}^{{}^c}}F^{\begin{scriptsize}Y_1\end{scriptsize}}_1\right]$ & $G^{(\Sigma^0)}_u\left[q'^2 \dfrac{F^{\begin{scriptsize}Y_2\end{scriptsize}}_2}{M} - {\bm{\Sigma}_{M_2}^{{}^c}}F^{\begin{scriptsize}Y_2\end{scriptsize}}_1\right]$ & $-$ \\ 
\hline
$\tilde{D}_2$ & $-G^{(\Lambda)}_u\dfrac{F^{\begin{scriptsize}Y_1\end{scriptsize}}_2}{2M}$ & $-G^{(\Sigma^0)}_u\dfrac{F^{\begin{scriptsize}Y_2\end{scriptsize}}_2}{2M}$ & $-$ \\
\hline
$\tilde{D}_3$ & $-$ & $-$ & $-$ \\
\hline
$\tilde{D}_4$ & $G^{(\Lambda)}_u\dfrac{F^{\begin{scriptsize}Y_1\end{scriptsize}}_2}{M}$ & $G^{(\Sigma^0)}_u\dfrac{F^{\begin{scriptsize}Y_2\end{scriptsize}}_2}{M}$ & $-$ \\ 
\hline
$\tilde{D}_5$ &  $-G^{(\Lambda)}_uG^{\begin{scriptsize}Y_1\end{scriptsize}}_A$ & $-G^{(\Sigma^0)}_uG^{\begin{scriptsize}Y_2\end{scriptsize}}_A$ & $-$ \\ 
\hline
$\tilde{D}_6$ & $-G^{(\Lambda)}_u\left[F^{\begin{scriptsize}Y_1\end{scriptsize}}_1 + \bm{\Delta}M_2\dfrac{F^{\begin{scriptsize}Y_1\end{scriptsize}}_2}{2M} \right]$ & $-G^{(\Sigma^0)}_u\left[ F^{\begin{scriptsize}Y_2\end{scriptsize}}_1 + \bm{\Delta}M_1\dfrac{F^{\begin{scriptsize}Y_2\end{scriptsize}}_2}{2M} \right]$ & $-$\\
\end{tabular}
\end{center}
\end{ruledtabular}
\end{footnotesize}
\label{tab:app3_0}
\end{table}

\begin{table}[h]
\caption{Extraction of the unknown amplitudes for the CC2 reaction process.}
\centering
\begin{footnotesize}
\begin{ruledtabular}
\begin{center}
\begin{tabular}{cccc}
Amplitude & $s$ channel ($p$) & $u$ channel ($\Lambda$) & $u$ channel ($\Sigma^0$) \\ 
\hline 
$\tilde{A}_1$ & $-$ & $2{G'^{(\Lambda)}_u}^{\footnote{\begin{scriptsize}$G'_s = g_{\text{K}^0 \Sigma^+ \text{p}} (s - M^2_N)^{-1},\,\,\,G'^{(\Lambda)}_{u} = g_{\text{K}^0 \Lambda \text{n}} (u - M^2_{\Lambda})^{-1},\,\,\,G'^{(\Sigma^0)}_{u} = g_{\text{K}^0 \Sigma^0 \text{n}} (u - M^2_{\Sigma^0})^{-1}
$.  \end{scriptsize}}}{G^{\begin{scriptsize}Y_1\end{scriptsize}}_A}^{\footnote{\begin{scriptsize}$F^{N'}_i(Q^2) = f^{p}_i(Q^2) - f^{n}_i(Q^2),\,G^{N'}_A(Q^2) = g_A(Q^2);\,F^{Y'_1}_i(Q^2) = -\sqrt{\frac{3}{2}}f^{n}_i(Q^2),G^{Y'_1}_A(Q^2) = \sqrt{\frac{2}{3}}\frac{D}{F + D} g_A(Q^2);\\F^{Y'_2}_i(Q^2) =  -\frac{1}{\sqrt{2}}(2f^{p}_i(Q^2)) + f^{n}_i(Q^2),\,G^{Y'_2}_A(Q^2) = -\sqrt{2}\frac{F}{F + D}g_A(Q^2)$.\end{scriptsize} }}$ & $2{G'^{(\Sigma^0)}_u}^{{}^a}{G^{\begin{scriptsize}Y'_2\end{scriptsize}}_A}^{{}^b}$  \\
\hline
$\tilde{A}_2$ & $-2{G'_s}^{{}^a}{G^{\begin{scriptsize}N'\end{scriptsize}}_A}^{{}^b}$ & $-$ & $-$ \\
\hline
$\tilde{A}_3$ & $-$ & $-2G'^{(\Lambda)}_uG^{\begin{scriptsize}Y'_1\end{scriptsize}}_A$ & $- 2G'^{(\Sigma^0)}_uG^{\begin{scriptsize}Y'_2\end{scriptsize}}_A$ \\
\hline
$\tilde{A}_4$ & $-$ & $-$ & $-$ \\ 
\hline
$\tilde{B}_1$ & $-G'_s\dfrac{M_N}{M}{F^{\begin{scriptsize}N'\end{scriptsize}}_2}^{{}^b}$ & $G'^{(\Lambda)}_u\left[2{F^{\begin{scriptsize}Y'_1\end{scriptsize}}_1}^{{}^b} + {\bm{\Delta}M_2}^{\footnote{\begin{scriptsize}$\bm{\Delta}M_1 = M_{\Sigma^+} - M_{\Sigma^0},\,\,\bm{\Delta}M_2 = M_{\Sigma^+} - M_{\Lambda},\,\,\bm{\Sigma}_{M_2} = M_{\Sigma^+} + M_{\Sigma^0},\,\,\bm{\Sigma}_{M_3} = M_{\Sigma^+} + M_{\Lambda} $.\end{scriptsize} }}\dfrac{{F^{\begin{scriptsize}Y'_1\end{scriptsize}}_2}^{{}^b}}{2M} \right] $ & $G'^{(\Sigma^0)}_u\left[ 2{F^{\begin{scriptsize}Y'_2\end{scriptsize}}_1}^{{}^b} + {\bm{\Delta}M_1}^{{}^c}\dfrac{{F^{\begin{scriptsize}Y'_2\end{scriptsize}}_2}^{{}^b}}{2M} \right]$ \\ 
\hline
$\tilde{B}_2$ & $2G'_s{F^{\begin{scriptsize}N'\end{scriptsize}}_1}^{{}^b}$ & $-$ & $-$ \\
\hline  
$\tilde{B}_3$ & $-$ & $-2G'^{(\Lambda)}_uF^{\begin{scriptsize}Y'_1\end{scriptsize}}_1$ & $-2G'^{(\Sigma^0)}_uF^{\begin{scriptsize}Y'_2\end{scriptsize}}_1$ \\
\hline
$\tilde{B}_4$ & $-$ & $-$ & $-$ \\ 
\hline
$\tilde{C}_1$ & $2M_NG'_s G^{\begin{scriptsize}N'\end{scriptsize}}_A$ & $G'^{(\Lambda)}_u {\bm{\Delta}M_2}G^{\begin{scriptsize}Y'_1\end{scriptsize}}_A$ & $G'^{(\Sigma^0)}_u {\bm{\Delta}M_1}G^{\begin{scriptsize}Y'_2\end{scriptsize}}_A$ \\
\hline
$\tilde{C}_2$ & $-$ & $-$ & $-$ \\ 
\hline
$\tilde{C}_3$ & $-$  & $-$ & $-$ \\ 
\hline
$\tilde{C}_4$ & $-$ & $-$ & $-$ \\ 
\hline
$\tilde{D}_1$ & $G'_s\left( \dfrac{q^2}{2} + {p_1 \cdot q}\right)\dfrac{F^{\begin{scriptsize}N'\end{scriptsize}}_2}{M}$ &  $G'^{(\Lambda)}_u\left[ {q'^2}^{\footnote{$q'^2 = ({q^2}/{2} - q \cdot p'_2 )$.}} \dfrac{F^{\begin{scriptsize}Y'_1\end{scriptsize}}_2}{M} - \bm{\Sigma}_{M_3}^{{}^{\,\,c}}F^{\begin{scriptsize}Y'_1\end{scriptsize}}_1\right]$ & $G'^{(\Sigma^0)}_u\left[q'^2 \dfrac{F^{\begin{scriptsize}Y'_2\end{scriptsize}}_2}{M} - \bm{\Sigma}_{M_2}^{{}^{\,\,c}}F^{\begin{scriptsize}Y'_2\end{scriptsize}}_1\right]$\\ 
\hline
$\tilde{D}_2$ & $-G'_s\dfrac{F^{\begin{scriptsize}N'\end{scriptsize}}_2}{2M}$ & $-G'^{(\Lambda)}_u\dfrac{F^{\begin{scriptsize}Y'_1\end{scriptsize}}_2}{2M}$ & $-G'^{(\Sigma^0)}_u\dfrac{F^{\begin{scriptsize}Y'_2\end{scriptsize}}_2}{2M}$ \\
\hline
$\tilde{D}_3$ & $-G'_s\dfrac{F^{\begin{scriptsize}N'\end{scriptsize}}_2}{M}$ & $-$ & $-$ \\
\hline
$\tilde{D}_4$ & $-$ & $G'^{(\Lambda)}_u\dfrac{F^{\begin{scriptsize}Y'_1\end{scriptsize}}_2}{M}$ & $G'^{(\Sigma^0)}_u\dfrac{F^{\begin{scriptsize}Y'_2\end{scriptsize}}_2}{M}$ \\ 
\hline
$\tilde{D}_5$ & $-G'_sG^{\begin{scriptsize}N'\end{scriptsize}}_A$ &  $-G'^{(\Lambda)}_uG^{\begin{scriptsize}Y'_1\end{scriptsize}}_A$ & $-G'^{(\Sigma^0)}_uG^{\begin{scriptsize}Y'_2\end{scriptsize}}_A$ \\ 
\hline
$\tilde{D}_6$ & $G'_s(F^{\begin{scriptsize}N'\end{scriptsize}}_1 + \dfrac{M_N}{M}F^{\begin{scriptsize}N'\end{scriptsize}}_2)$ & $-G'^{(\Lambda)}_u\left[F^{\begin{scriptsize}Y'_1\end{scriptsize}}_1 + {\bm{\Delta}M_2}\dfrac{F^{\begin{scriptsize}Y'_1\end{scriptsize}}_2}{2M} \right]$ & $-G'^{(\Sigma^0)}_u\left[ F^{\begin{scriptsize}Y'_2\end{scriptsize}}_1 + {\bm{\Delta}M_1}\dfrac{F^{\begin{scriptsize}Y'_2\end{scriptsize}}_2}{2M} \right]$ \\
\end{tabular}
\end{center}
\end{ruledtabular}
\end{footnotesize}
\label{tab:app3_1}
\end{table}
\begin{table}[h]
\caption{Extraction of the unknown amplitudes for the CC3 reaction process.}
\centering
\begin{footnotesize}
\begin{ruledtabular}
\begin{center}
\begin{tabular}{cccc}
Amplitude & $s$ channel ($p$) & $u$ channel ($\Sigma^-$) & $t$ channel ($\text{K}^0$)\\ 
\hline 
$\tilde{A}_1$ & $-$ & $2\tilde{G}_u^{\footnote{\begin{scriptsize}$\tilde{G}_s = g_{\text{K}^+ \Sigma^0 \text{p}}(s - M^2_N)^{-1},\,\,\, \tilde{G}_u = g_{\text{K}^{+} \Sigma^{-} \text{n}}(u - M^2_{\Sigma^-})^{-1} ;\,\,\,\tilde{G}_t = g_{\text{K}^0\Sigma^0 \text{n}}(t - M^2_{K^0})^{-1}$.\end{scriptsize} }}\tilde{G}_A^{{\begin{scriptsize}Y\end{scriptsize}}^{\footnote{\begin{scriptsize}$\tilde{F}^{N}_i(Q^2) = f^{p}_i(Q^2) - f^{n}_i(Q^2),\,\tilde{G}^{N}_A(Q^2) = g_A(Q^2); \,\tilde{F}^{Y}_i(Q^2) = \frac{1}{\sqrt{2}}(2f^{p}_i(Q^2) + f^{n}_i(Q^2)),\\ \tilde{G}^{Y}_A(Q^2) = \sqrt{2}\frac{F}{F + D}g_A(Q^2)$.\end{scriptsize} }}}$ & $-$ \\
\hline
$\tilde{A}_2$ & $-2\tilde{G}_s^{{}^a}\tilde{G}_A^{\begin{scriptsize}N\end{scriptsize}^{{}^b}}$ & $-$ & $-$ \\
\hline
$\tilde{A}_3$ & $-$ & $-2\tilde{G}_u\tilde{G}^{\begin{scriptsize}Y\end{scriptsize}}_A$ & $-$\\
\hline
$\tilde{A}_4$ & $-$ & $-$ & $-$\\ 
\hline
$\tilde{B}_1$ & $-\tilde{G}_s\dfrac{M_N}{M}\tilde{F}^{{\begin{scriptsize}N\end{scriptsize}}^{{}^b}}_2$ & $\tilde{G}_u\left[ 2\tilde{F}^{{\begin{scriptsize}Y\end{scriptsize}}^{{}^b}}_1 + \bm{\Delta}_{M_3}^{\footnote{\begin{scriptsize}$\bm{\Delta}_{M_3} =  M_{\Sigma0} -  M_{\Sigma^{-}},\,\,\,\bm{\Sigma}_{M4} = M_{\Sigma0} + M_{\Sigma^{-}}$.\end{scriptsize} }}\dfrac{\tilde{F}^{{\begin{scriptsize}Y\end{scriptsize}}^{{}^b}}_2}{2M}  \right]$ & $\tilde{G}_t^{{}^a}F_{K^{+,\,0}}$ \\ 
\hline
$\tilde{B}_2$ & $2\tilde{G}_s\tilde{F}^{{\begin{scriptsize}N\end{scriptsize}}^{{}^b}}_1$ & $-$ & $2\tilde{G}_tF_{K^{+,\,0}}$\\
\hline  
$\tilde{B}_3$ & $-$ & $-2\tilde{G}_u\tilde{F}^{\begin{scriptsize}Y\end{scriptsize}}_1$ & $-2\tilde{G}_tF_{K^{+,\,0}}$ \\
\hline
$\tilde{B}_4$ & $-$ & $-$ & $-$\\ 
\hline
$\tilde{C}_1$ & $2M_N\tilde{G}_s \tilde{G}^{\begin{scriptsize}N\end{scriptsize}}_A$ & $\tilde{G}_u\bm{\Delta}_{M_3}\tilde{G}^{\begin{scriptsize}Y\end{scriptsize}}_A$ &$-$ \\
\hline
$\tilde{C}_2$ & $-$ & $-$ & $-$ \\ 
\hline
$\tilde{C}_3$ & $-$ & $-$ & $-$ \\ 
\hline
$\tilde{C}_4$ & $-$ & $-$ & $-$ \\ 
\hline
$\tilde{D}_1$ & $\tilde{G}_s\left( \dfrac{q^2}{2} + {p_1 \cdot q}\right)\dfrac{\tilde{F}^{\begin{scriptsize}N\end{scriptsize}}_2}{M}$ & $-\tilde{G}_u\left[ \bm{\Sigma}_{M4}^{{}^{\,\,c}}\tilde{F}^{\begin{scriptsize}Y\end{scriptsize}}_1 + \left(q \cdot {p'}_2 - \dfrac{q^2}{2}\right) \dfrac{\tilde{F}^{\begin{scriptsize}Y\end{scriptsize}}_2}{M}\right]$ & $-$ \\ 
\hline
$\tilde{D}_2$ & $-\tilde{G}_s\dfrac{\tilde{F}^{\begin{scriptsize}N\end{scriptsize}}_2}{2M}$ & $-\tilde{G}_u\dfrac{\tilde{F}^{\begin{scriptsize}Y\end{scriptsize}}_2}{2M}$ & $-$ \\
\hline
$\tilde{D}_3$ & $-\tilde{G}_s\dfrac{\tilde{F}^{\begin{scriptsize}N\end{scriptsize}}_2}{M}$ & $-$ & $-$ \\
\hline
$\tilde{D}_4$ & $-$ & $\tilde{G}_u\dfrac{\tilde{F}^{\begin{scriptsize}Y\end{scriptsize}}_2}{M}$ & $-$ \\ 
\hline
$\tilde{D}_5$ &  $-\tilde{G}_s\tilde{G}^{\begin{scriptsize}N\end{scriptsize}}_A$ & $-\tilde{G}_u\tilde{G}^{\begin{scriptsize}Y\end{scriptsize}}_A$ & $-$ \\ 
\hline
$\tilde{D}_6$ & $\tilde{G}_s(\tilde{F}^{\begin{scriptsize}N\end{scriptsize}}_1 + \dfrac{M_N}{M}\tilde{F}^{\begin{scriptsize}N\end{scriptsize}}_2) $ & $-\tilde{G}_u\left[ \tilde{F}^{\begin{scriptsize}Y\end{scriptsize}}_1 + \left(\bm{\Delta}_{M_3}\right)\dfrac{\tilde{F}^{\begin{scriptsize}Y\end{scriptsize}}_2}{2M} \right]$ & $-$\\
\end{tabular}
\end{center}
\end{ruledtabular}
\end{footnotesize}
\label{tab:app3_2}
\end{table}

\begin{table}[h]
\caption{Extraction of the unknown amplitudes for the CC4 reaction process.}
\centering
 \begin{footnotesize}
\begin{ruledtabular}
\begin{center}
% use packages: array
\begin{tabular}{cccc}
Amplitudes & $s$ channel & $u$ channel & $t$ channel\\ 
\hline 
$\tilde{A}_1$ & $-$ & $2 G_u^{\footnote[1]{\begin{scriptsize}$G_s = g_{\text{K}^+\Lambda \text{p}}({s - M_{N}})^{-1}$,\,\,\,$G_t = g_{K^0\Lambda n}({t - M^2_{K^0}})^{-1}$,\,\,\,$G_u = g_{K^+\Sigma^{-} n}({u - M_{\Sigma^{-}}})^{-1}$.\end{scriptsize} }}{G^{\begin{scriptsize}Y\end{scriptsize}}_A}^{\footnote[2]{\begin{scriptsize}${F}^{N}_i(Q^2) = f^{p}_i(Q^2) - f^{n}_i(Q^2),\,{G}^{N}_A(Q^2) = g_A(Q^2);\,{F}^{Y}_i(Q^2) = -\sqrt{\frac{3}{2}}f^{n}_i(Q^2),\,{G}^{Y}_A(Q^2) = \sqrt{\frac{2}{3}}\frac{D}{F + D}g_A(Q^2).$\end{scriptsize} }}$ & $-$ \\
\hline
$\tilde{A}_2$ & $-2G_s^{{}^a} {G^{\begin{scriptsize}N\end{scriptsize}}_A}^{{}^b}$ & $-$ & $-$ \\
\hline
$\tilde{A}_3$ & $-$ & $- 2G_uG^{\begin{scriptsize}Y\end{scriptsize}}_A$ & $-$\\
\hline
$\tilde{A}_4$ & $-$ & $-$ & $-$\\ 
\hline
$\tilde{B}_1$ & $-G_s\dfrac{M_N}{M}{F^{\begin{scriptsize}N\end{scriptsize}}_2}^{{}^b}$ & $G_u\left[ 2{F^{\begin{scriptsize}Y\end{scriptsize}}_1}^{{}^b} + \bm{\Delta}_M^{\footnote[3]{\begin{scriptsize}$\bm{\Delta}_M = M_{\Lambda} -  M_{\Sigma^{-}}$,\,\,\,$\bm{\Sigma}_{M_1} = M_{\Lambda} + M_{\Sigma^{-}}$.\end{scriptsize} }}\dfrac{{F^{\begin{scriptsize}Y\end{scriptsize}}_2}^{{}^b}}{2M}  \right]$ & $G_t^{{}^a}F_{K^{+,\,0}}$ \\ 
\hline
$\tilde{B}_2$ & $2G_s{F^{\begin{scriptsize}N\end{scriptsize}}_1}^{{}^b}$ & $-$ & $2G_tF_{K^{+,\,0}}$\\
\hline  
$\tilde{B}_3$ & $-$ & $-2G_uF^{\begin{scriptsize}Y\end{scriptsize}}_1$ & $-2G_tF_{K^{+,\,0}}$ \\
\hline
$\tilde{B}_4$ & $-$ & $-$ & $-$\\ 
\hline
$\tilde{C}_1$ & $2M_NG_s G^{\begin{scriptsize}N\end{scriptsize}}_A$ & $G_u\bm{\Delta}_MG^{\begin{scriptsize}Y\end{scriptsize}}_A$ &$-$ \\
\hline
$\tilde{C}_2$ & $-$ & $-$ & $-$ \\ 
\hline
$\tilde{C}_3$ & $-$ & $-$ & $-$ \\ 
\hline
$\tilde{C}_4$ & $-$ & $-$ & $-$ \\ 
\hline
$\tilde{D}_1$ & $G_s\left( \dfrac{q^2}{2} + {p_1 \cdot q}\right)\dfrac{F^{\begin{scriptsize}N\end{scriptsize}}_2}{M}$ & $-G_u\left[\bm{\Sigma}_{M_1}^{{}^{\,\,c}} F^{\begin{scriptsize}Y\end{scriptsize}}_1 + \left(q \cdot {p'}_2 - \dfrac{q^2}{2}\right) \dfrac{F^{\begin{scriptsize}Y\end{scriptsize}}_2}{M}\right]$ & $-$ \\ 
\hline
$\tilde{D}_2$ & $-G_s\dfrac{F^{\begin{scriptsize}N\end{scriptsize}}_2}{2M}$ & $-G_u\dfrac{F^{\begin{scriptsize}Y\end{scriptsize}}_2}{2M}$ & $-$ \\
\hline
$\tilde{D}_3$ & $-G_s\dfrac{F^{\begin{scriptsize}N\end{scriptsize}}_2}{M}$ & $-$ & $-$ \\
\hline
$\tilde{D}_4$ & $-$ & $G_u\dfrac{F^{\begin{scriptsize}Y\end{scriptsize}}_2}{M}$ & $-$ \\ 
\hline
$\tilde{D}_5$ &  $-G_sG^{\begin{scriptsize}N\end{scriptsize}}_A$ & $-G_uG^{\begin{scriptsize}Y\end{scriptsize}}_A$ & $-$ \\ 
\hline
$\tilde{D}_6$ & $G_s(F^{\begin{scriptsize}N\end{scriptsize}}_1 + \dfrac{M_N}{M}F^{\begin{scriptsize}N\end{scriptsize}}_2) $ & $-G_u\left[ F^{\begin{scriptsize}Y\end{scriptsize}}_1 + \bm{\Delta}_M\dfrac{F^{\begin{scriptsize}Y\end{scriptsize}}_2}{2M} \right]$ & $-$\\
\end{tabular}
\end{center}
\end{ruledtabular}
\end{footnotesize}
\label{tab:app3_3}
\end{table}

\clearpage
\bibliography{adera_aps}

\end{document}